\documentclass[aps,prx,reprint,nofootinbib,twocolumn,superscriptaddress,showpacs,showkeys,longbibliography]{revtex4-1}

\usepackage{eurosym}
\usepackage{amsmath,amssymb,amstext}
\usepackage[usenames,dvipsnames]{color}
\usepackage{graphicx}
\usepackage{braket}
\usepackage{natbib}
\usepackage{comment}
\usepackage{dcolumn}
\usepackage[english]{babel}
\usepackage{wasysym}
\usepackage[colorlinks,bookmarks=false,citecolor=blue,linkcolor=red,urlcolor=blue]{hyperref}

\begin{document}

\title{Majoranas in Noisy Kitaev Wires}

\author{Ying Hu}
\affiliation{Institute for Quantum Optics and Quantum Information of the Austrian Academy of Sciences, A-6020 Innsbruck, Austria}

\author{Zi Cai}
\affiliation{Institute for Quantum Optics and Quantum Information of the Austrian Academy of Sciences, A-6020 Innsbruck, Austria}

\author{Mikhail A. Baranov}

\affiliation{Institute for Quantum Optics and Quantum Information of the Austrian Academy of Sciences, A-6020 Innsbruck, Austria}
\affiliation{NRC Kurchatov Institute, Kurchatov Square 1, 123182 Moscow, Russia}

\author{Peter Zoller}

\affiliation{Institute for Quantum Optics and Quantum Information of the Austrian Academy of Sciences, A-6020 Innsbruck, Austria}
\affiliation{Institute for Theoretical Physics, University of Innsbruck, A-6020 Innsbruck, Austria}

\begin{abstract}
Robustness of edge states and non-Abelian excitations of topological states
of matter promises quantum memory and quantum processing, which is naturally
immune against microscopic imperfections such as static disorder. However,
topological properties will not in general protect quantum system from
time-dependent disorder or noise. Here we take the example of a network of
Kitaev wires with Majorana edge modes storing qubits to investigate the
effects of classical noise in the crossover from the quasi-static to the
fast fluctuation regime. We present detailed results for the Majorana edge
correlations, and fidelity of braiding operations for both global and local
noise sources preserving parity symmetry, such as random chemical potentials
and phase fluctuations. While in general noise will induce heating and
dephasing, we identify examples of long-lived quantum correlations in
presence of fast noise due to motional narrowing, where external noise drives the
system rapidly between the topological and non-topological phases.
\end{abstract}

\date{\today}
\maketitle


\section{Introduction\label{sec:intro}}

At present there is significant interest and ongoing effort in realizing and detecting
topological phases of quantum matter in the laboratory \cite
{Konig2007,Hsieh2008,Chen2009,Bloch2013,Goldman2013,Esslinger2014,Schneider2015,Fallani2015,Spielman2015}. These efforts are driven by both foundational aspects of our understanding
of quantum ordered phases in many-body systems beyond the Landau paradigm of
local order parameters (see e.g. Refs. \cite{Wen2004,Hasan2010,Qi2011}), and in particular by
the promises to use the intrinsic robustness of topological properties
against imperfections in quantum information processing \cite
{Kitaev2003,DasSarma2005,Nayak2008,Pachos2012,AliceaStern2014}. An example is provided by
Kitaev's quantum wire \cite{Kitaev2001} supporting a pair of Majorana edge
modes - Majorana fermions, which show non-Abelian exchange statistics under
braiding \cite{Ivanov2001,Alicea2011} and represent a topologically
protected non-local zero-energy fermion. In a wire network, these properties
can be used to create topologically protected qubits and gate operations 
\cite{Alicea2011}. The quest to demonstrate Majorana fermions and their
non-Abelian properties is presently an outstanding challenge in quantum
physics \cite{Wilczek2009,Alicea2013_review,Beenakker2013,DasSarma2014}, and
is the focus of a significant effort involving systems from hybrid
nano-wires \cite{Sau2010,Lutchyn2010,Alicea2010,Oreg2010,Halperin2012,Romito2012,Mourik2012,Deng2012,Rokhinson2012,Das2012,Churchill2013,Finck2013,NadjPerge2014,Alicea2011}
to cold atom setups \cite{Jiang2011,Nascimbene2013,Diehl2011,Sato2009,Kraus2012,Kraus2013pair,Buhler2014,Kraus2013,Laflamme2014}. First evidence for Majorana edge modes has been reported in recent
experiments \cite{Mourik2012,Deng2012,Rokhinson2012,Das2012,Churchill2013,Finck2013,NadjPerge2014}.

\begin{figure}[tb]
\centering
\includegraphics[width= 1.0\columnwidth]{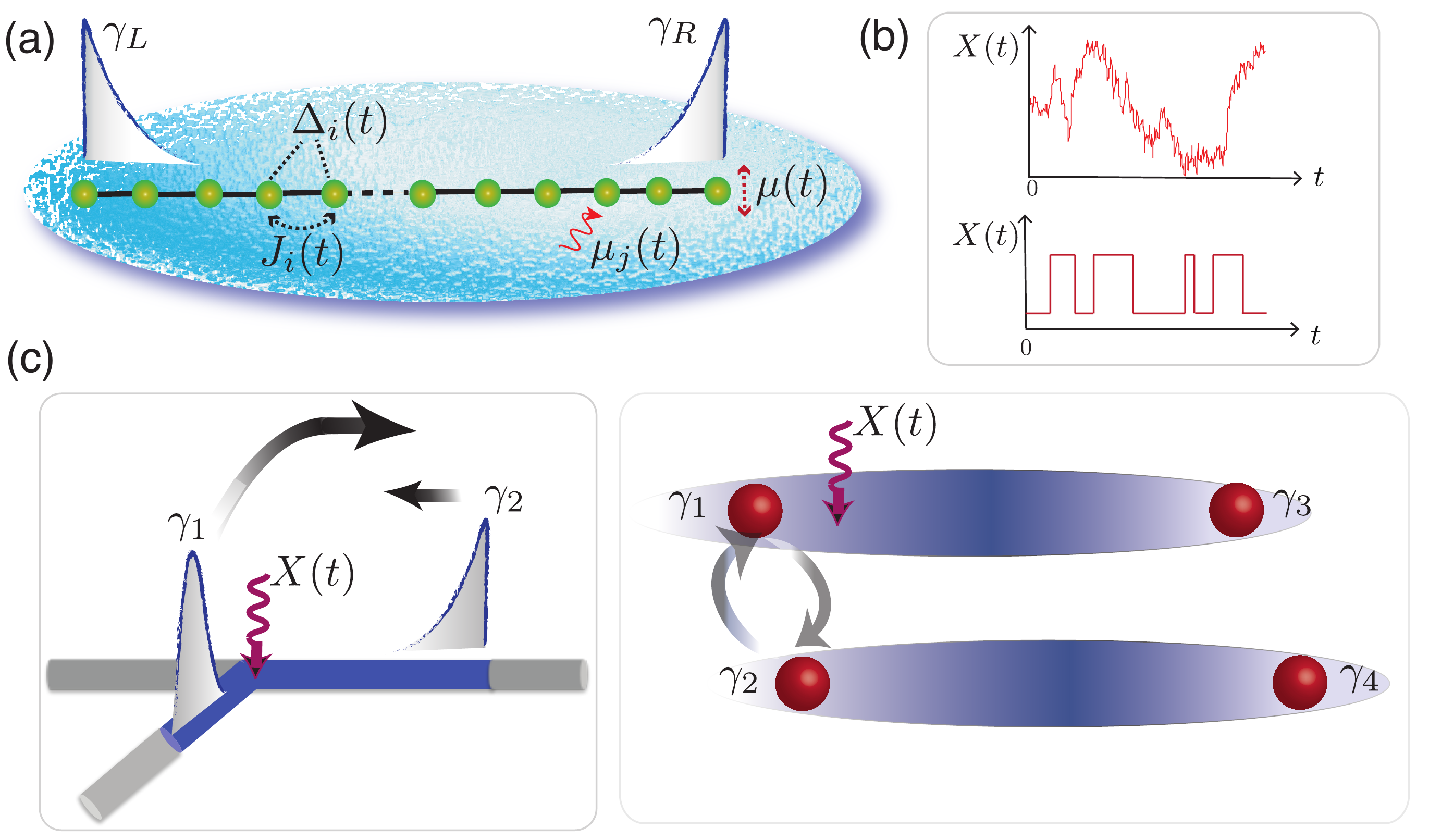} 
\caption{Schematic of Majorana edge modes in noisy Kitaev wires and braiding operations in presence of colored Markovian noise.  ({a}) A Kitaev wire in the topological phase supports two Majorana edge modes $\gamma_{L/R}$. In realistic implementations, a noisy Kitaev Hamiltonian (\ref{Kitaev}) containing stochastic parameters $X(t)\equiv\{J_j(t),\Delta_j(t),\mu_j(t)\}$ arises, due to a coupling to a classical environment (blue background). The fluctuations can be either global [e.g. $\mu(t)$], or local [e.g. $\mu_j(t)$], depending on the physical sources of noise in specific implementations.  ({b}) Two typical examples of colored Markovian noise $X(t)$:  the colored Gaussian noise (upper panel) and the two-state telegraph noise (lower panel). (c) Braiding Majoranas (say, $\gamma_1$ and $\gamma_2$) on a noisy T-junction in solid-state setting (left panel), and in a noisy atomic wire network in the optical lattice setup (right panel).}\label{Fig:schematics}
\end{figure}

However, while the promise of topological protection of quantum states from
microscopic imperfections may hold for static disorder (see e.g. Refs. \cite{Nayak2008,Pachos2012}), recent theoretical studies have concluded that
Majorana qubits and braiding can be seriously affected by coupling to an
environment \cite{Goldstein2011,Budich2012,Raninis2012,Pedrocchi2015,Hu2014}, as will be the case in any realistic experimental scenario. The protection
of Majorana modes in the Kitaev wire is related to protection of fermion
parity, and quantum correlations between Majorana states will be rapidly
destroyed by injection or removal of quasiparticles \cite{Raninis2012}. Even
the coupling to a finite temperature bosonic bath, which preserves particle
parity, is predicted to result in unavoidable losses of coherence and errors 
\cite{Pedrocchi2015}. Nevertheless, as we will show in this paper, it is
possible to identify examples with long-lived quantum correlations between
Majorana states in the presence of noise. In particular, we will be
interested in the effects of local and global noise representing a
parity-preserving coupling to an environment, which we model as a classical
stochastic process. The case of \textit{local} noise is representative of a
two-level fluctuator in a solid state realization of a Kitaev wire\footnote{Decoherence of a superconducting qubit due to coupling to a single two-state  charge-fluctuator or to a set of them with distributed switching rates and couplings to the qubit, was considered, for example in Refs. \cite{Grishin2005,Galperin2006,Cheng2008}.}, while 
\textit{global} fluctuations can result, for example, from laser light
fluctuations in cold-atom experiments.

Our goal is thus to study the effects of noise on Majorana correlations and
braiding operations, in a regime ranging from quasi-static disorder, all the
way to the limit of fast fluctuations, i.e. where the noise correlation time
is much shorter than the relevant system time scales. Although coupling to
classical noise will eventually always lead to dephasing and heating,
dephasing can be suppressed in the fast fluctuation limit, even when the
system is driven by the noise e.g. between topological and non-topological
phases. This effect of noise suppression with decreasing correlation time is familiar from atomic physics as motional narrowing. There, increasing the collision rate between atoms can result in a narrowing of the spectral lines  \cite{CohenTannoudji1977}, and in our context in an increased coherence time of Majorana correlations. In such cases, we will also determine the optimal conditions for
braiding time scales - as a trade off between the requirement of
adiabaticity of braiding, and decoherence time scales.

The emphasis and value of the present work is on {\em exactly solvable} models quantum many-body dynamics of Majorana correlations and braiding operations in the presence of {\em colored Markovian noise} sources, as exemplified by telegraph noise,  $n$-state jump models, or colored Gaussian noise. While in a solid state context this should be understood as phenomenological models of noise describing imperfections like local fluctuators, we note that noise in cold atom experiments can be engineered, as in recent studies of Anderson localization with (static) random optical potentials \cite{Giacomo2008,Billy2008}. Atomic realizations of Majorana fermions may thus serve as an ideal platform to study the effect of {\em time dependent disorder} in a controlled setting by appropriate modulation of the laser beams to mimic various noise sources, and can thus provide a direct experimental counterpart to the present theoretical study.

The paper is organized as follows. In Sec. \ref{sec:Kitaev}, we briefly
describe the model Hamiltonian of a noisy Kitaev wire. Then, in Sec. \ref{sec:marginal}, we develop techniques which allow non-perturbative solutions
of the many-body quantum dynamics for colored Markovian noise with arbitrary
correlation time. Based on these techniques, Sec. \ref{sec:global} presents a
study on the Majorana edge correlations and heating dynamics for a global
noise, which stochastically drives the system e.g. across the boundary
between the topological and non-topological phases. The case of local noise
is investigated in Sec. \ref{sec:local}. Thereafter, in Sec. \ref{sec:move},
we study the effect of colored noise on Majorana transport, and discuss
optimal conditions for Majorana manipulations to obtain the best fidelity at
a given noise. The braiding dynamics on a noisy wire network is analyzed in
Sec. \ref{sec:braid}, based on the noisy T-junction architecture and
cold-atom setup, respectively. The paper closes with a summary and outlook
in Sec. \ref{sec:conc}.

\section{Noisy Kitaev wire}

\label{sec:Kitaev}

Our goal is to study the dynamics of Majorana edge modes of the Kitaev wire
in the presence of noise (see Fig. \ref{Fig:schematics}). The relevant
Hamiltonian is 
\begin{eqnarray}
H[X(t)]&=&\sum_{j=1}^{N-1}\left[-J_{j}(t)a_{j}^{\dag }a_{j+1}+\Delta
_{j}(t)a_{j}a_{j+1}+\text{h.c.}\right]  \notag \\
&-&\sum_{j=1}^{N}\mu _{j}(t)a_{j}^{\dag}a_{j},  \label{Kitaev}
\end{eqnarray}
where $a_{j}$ and $a_{j}^{\dag}$ are the operator of spinless fermions on a
finite chain of $N$ sites. Here $J_{j}(t)$ is the hopping amplitude on the
lattice, $\Delta_{j}(t)$ is the pairing parameter and $\mu_{j}(t)$ is the
local chemical potential. We assume that these parameters fluctuate in time
according to a given noise model $X(t)\equiv\{J_{j}(t),\Delta_{j}(t),\mu_{j}(t)\}$, which is motivated by a particular physical noise source related
to a specific implementation. In hybrid nano-wires the above Hamiltonian (or
its continuous version) arises from a combination of spin-orbit coupling of
electrons in the presence of a magnetic field, and the coupling to an $s$-wave
superconductor \cite{Sau2010,Lutchyn2010,Alicea2010,Oreg2010,Mourik2012,Deng2012,Rokhinson2012,Das2012,Churchill2013,Finck2013,NadjPerge2014}. Thus the effect of a two-level fluctuator, for example, can be represented
by a fluctuating \emph{local} chemical potential $\mu_{j}(t)$ on a given
lattice site. On the other hand, a realization of the Kitaev wire with cold
fermionic atoms in a 1D optical lattice results from a laser induced
coupling to a molecular Bose-Einstein condensate, where a molecule is
dissociated into a pair of fermions in the wire, thus realizing the pairing
term $\Delta_{j}$ \cite{Jiang2011,Nascimbene2013,Hu2014}. Frequency
fluctuations of the laser light can be understood as fluctuations of the
laser detuning, or as a \emph{global} fluctuating chemical potential $\mu_{j}(t)\equiv\mu(t)$.

Before investigating the dynamics of the noisy wire (\ref{Kitaev}), let us
briefly describe as a reference the properties of a noise-free Kitaev's
quantum wire \cite{Kitaev2001}, where $J_{j}(t)=J$, $\Delta_{j}(t)=\Delta$,
and $\mu_{j}(t)=\mu$. When $|\mu|<2J$ and $\Delta\neq 0$, the wire is in
the topological phase characterized by a gapped energy spectrum in the bulk
and by a pair of robust Majorana edge modes $\gamma_{L}=\gamma_{L}^{\dag}$
and $\gamma_{R}=\gamma_{R}^{\dag}$ of the form $\gamma
_{L/R}=\sum_jf_{L/R,j}c_{j}$ [here we use the Majorana representation $
c_{2j-1}=a_{j}+a_{j}^{\dag }$ and $c_{2j}=-i(a_{j}-a_{j}^{\dag})$ of the
operators] with $f_{L/R,j}$ being exponentially localized near the left ($L$) and right ($R$) edges with the localization length $l_{M}$. The pair of
Majorana edge modes represents a non-local fermionic \emph{zero-energy} mode 
$\alpha _{M}=(\gamma _{L}+i\gamma _{R})/2$, and the long-range correlations $
-i\langle\gamma_{L}\gamma_{R}\rangle$ between the Majorana
edge states, which are of interest for us, are directly related to the occupation of this mode, $-i\langle\gamma_{L}\gamma_{R}\rangle
=1-2\langle\alpha_{M}^{\dag}\alpha_{M}\rangle$. The
robustness of the correlations is therefore related to the conservation of
fermionic parity which distinguishes the two degenerate ground states with
empty and occupied mode $\alpha_{M}$. For $|\mu |>2J$, when the wire is in
the non-topological phase, there are no Majorana edge states and all
excitations are gapped.

The presence of noisy components in the parameters of the Hamiltonian (\ref{Kitaev}) results in random {\textquotedblleft shaking\textquotedblright}
of the system giving rise to changes in population of the $\alpha_{M}$-mode
(accompanied by creation of bulk excitations) and, therefore, to the decay
of the Majorana correlations. Large-amplitude global noise, say, in the
chemical potential, can also drive the system across the quantum phase
transition \textit{between different phases} -- topological and
non-topological ones, i.e. between the cases with two Majorana edge modes
and with none of them. Understanding the fate of Majorana edge modes, the
correlations between them, and their braiding in the noisy Kitaev wire (\ref{Kitaev}) involves the study of many-body non-equilibrium dynamics induced
by the noise, and we describe the developed techniques for treating such
problems in Sec. \ref{sec:marginal}.

\section{Quantum dynamics in colored Markovian noise}

\label{sec:marginal}

In dynamics of quantum systems external noise appears as stochastic
parameters in the Hamiltonian, as in Eq. (\ref{Kitaev}), and the
time-dependent density matrix equation becomes a multiplicative stochastic
differential equation \cite{Gardiner2010Stochastic,VanKampen1976}. Solving
for the averaged density matrix in presence of noise with arbitrary
correlation times, requires the development of non-perturbative techniques.
For fast fluctuations, when the noise correlation time is much shorter than
the system response time (white noise limit), a perturbative treatment, in form
of a lowest order cumulant expansion, results in a master equation for the
stochastically averaged density matrix \cite
{Gardiner2010Stochastic,VanKampen1976}. However, with increasing correlation
time, the system response will be sensitive to all higher order correlation
functions, and in the limit of infinite correlation time the static disorder
problem is recovered (as in Anderson \cite{Anderson1958}, and many-body
localization \cite{Huse2010}). In solving for quantum correlations and
braiding dynamics in the noisy Kitaev wire we will rely on Markovian models
of colored noise, which allow an exact solution of many-body dynamics for
arbitrary correlation time of the noise.

At the heart of our solution of quantum dynamics for colored Markovian noise
is the \textit{generalized master equation} for the \textit{marginal system
density operator}. In brief, we assume a Markov process $X(t)\equiv\{X_{\alpha}(t)\}$, and we denote by $H[X(t)]$ the associated system
Hamiltonian. Our goal is to solve the stochastic density matrix equation ($\hbar\equiv1$), $\partial_{t}\rho (t)=-i\Big[H[X(t)],\rho(t)\Big]$, with 
$\rho(t)$ the density operator, for the stochastic average $\langle\rho(t)\rangle_{s}$ with angular brackets denoting the noise average. Defining
a marginal density matrix 
\begin{equation}
\rho (X,t)=\left\langle\rho(t)\delta(X(t)-X)\right\rangle_{s},
\label{Marginal}
\end{equation}
we can derive the generalized master equation for the marginal density
matrix (see Ref. \cite{Peter1981} and App. \ref{App:derive}) 
\begin{equation}
\partial_{t}\rho \left(X,t\right)=\mathcal{L}\left(X\right)\rho \left(X,t\right)-i\left[ H\left(X\right),\rho \left(X,t\right)
\right] ,  \label{Master_density}
\end{equation}
with $X$ now a time independent variable. Here $\mathcal{L}(X)$ is generator of
our Markovian noise model, as appears in the differential Chapman-Kolmogorov
equation for the conditional density $P(X,t|X^{\prime},t^{\prime})$ \cite{Gardiner2010Stochastic} 
\begin{equation}
\partial_{t}P(X,t|X^{\prime },t^{\prime })=\mathcal{L}(X)P(X,t|X^{\prime},t^{\prime}).  \label{CK}
\end{equation}
Thus the operator $\mathcal{L}(X)$ provides a complete specification of our
noise model, and appears as a damping operator in the generalized master
equation (\ref{Master_density}). Solving (\ref{Master_density}) for $\langle\rho(t)\rangle_{s}=\int dX\rho(X,t)$ provides us with the desired
average. We refer to App. \ref{App:limit} for a discussion of various limits
of solving the above equation: this includes the regime of quasi-static
disorder, and the fast fluctuation limit (master equation limit). We
emphasize, however, that by solving (\ref{Master_density}) we obtain
solutions valid for \textit{arbitrary} correlation time.

The most general models for classical Markovian noises \cite{Gardiner2010Stochastic}  are described by diffusion processes with continuous noise
trajectory and by jump processes with $X(t)$ taking a set of discrete values $\{X_{m}\}$ ($m=0,...,N_{r}$). In the former case, $\mathcal{L}(X)$
corresponds to the Fokker-Planck operator, and a primary example is the
colored Gaussian noise (Ornstein-Uhlenbeck process), see upper panel of Fig. \ref{Fig:schematics}(b). In the latter $\mathcal{L}(X)$ reduces to a
matrix $\mathcal{L}_{mn}$ describing the jump-rates between different $X_{m}$, as in the case of multi-state telegraph noise, see lower panel of Fig. 
\ref{Fig:schematics}(b) for the two-state telegraph noise. While the
solution for a diffusion process can in principle be obtained in terms of the
eigenfunctions of the Fokker-Planck operator $\mathcal{L}(X)$ \cite
{Gardiner2010Stochastic, Peter1981}, we pursue here a much more convenient strategy by
discretizing it to a $N_{r}$-state jump model with large $N_{r}$
and properly chosen $\mathcal{L}_{mn}$ -- ``putting the
noise on a lattice". In this case the generalized master equation (\ref
{Master_density}) takes on the form of a set of coupled equations for marginal
density matrices $\rho(X_m,t)$
\begin{equation}
\!\!\!{\partial_t}\rho(X_m,t)=\sum_{n=0}^{N_r}\mathcal{L}_{mn}\rho(X_n,t)-i[H(X_m),\rho(X_m,t)], \label{Lattice}
\end{equation}
which - at least for low-dimensional processes $X(t)$ - is not
significantly of more effort to solve than the original non-stochastic version.

Below we will illustrate above techniques for the Markovian two-state jump
model [see lower panel of Fig. \ref{Fig:schematics}(b)], which represents the
simplest but relevant example capturing all essential features of noise
dynamics. It also allows a straightforward extension to multi-state jump
models, or to noise from several fluctuators, which for a large number in
the sense of the central limit theorem approaches colored Gaussian noise (see App. \ref{App:gaussian}). In addition, we will specialize
the above equations to the case of a quadratic many-body Hamiltonian, as
relevant for the Kitaev wire.

In two-state telegraph noise, the stochastic parameter $X(t)$ switches
randomly between two discrete values $a$ and $b$ with rate $\kappa$.
Equation (\ref{CK}) is now a rate equation for the probabilities $P(a,t)$
and $P(b,t)$ to be in state $a$ or $b$ respectively, 
\begin{equation}
\frac{d}{dt}\Big[
\begin{array}{c}
P(a,t) \\ 
P(b,t)
\end{array}
\Big]=\Big[
\begin{array}{cc}
-\kappa  & \kappa  \\ 
\kappa  & -\kappa 
\end{array}
\Big]\Big[
\begin{array}{c}
P(a,t) \\ 
P(b,t)
\end{array}
\Big]:=\mathcal{L}\Big[
\begin{array}{c}
P(a,t) \\ 
P(b,t)
\end{array}
\Big].  \label{Fluctuator}
\end{equation}
We have $\left\langle X\right\rangle _{s}=(a+b)/2$ and $\left\langle
X(t+\tau ),X(t)\right\rangle _{s}=\sigma ^{2}\exp\left( -|\tau |/\tau
_{c}\right) $ for the mean value and first order correlation function,
respectively, with $\tau _{c}=1/2\kappa $ the correlation time and the variance $
\sigma^{2}={(a-b)^{2}}/{4}$ (using the notation $\left\langle X,Y\right\rangle
_{s}\equiv\left\langle XY\right\rangle _{s}-\left\langle X\right\rangle_s
\left\langle Y\right\rangle_{s}$).

The dynamics of a Kitaev wire driven by a single telegraph noise can be
readily derived in the Majorana representation, where the quadratic
Hamiltonian (\ref{Kitaev}) can be recast as $H[X(t)]=(1/4)\sum_{il} h_{il}
[X(t)]c_ic_l$, with $h_{il}=-h_{li}$ being an asymmetric Hamiltonian matrix 
\cite{Kitaev2001}. In the Majorana basis, the key quantity capturing the
system dynamics is the covariance matrix $\langle{\Gamma(t)}\rangle_s=\text{
Tr}[\langle \rho(t)\rangle_{s}\hat{ \Gamma}]$, with $\hat{\Gamma}
_{il}=(i/2)[c_i,c_l]$. Thus from Eq. (\ref{Lattice}), we find for the
marginal densities 
\begin{eqnarray}
\frac{d}{d t}{\Gamma}(a,t)&=&-i[h(a),{\Gamma}(a,t)]-\kappa{\Gamma}
(a,t)+\kappa{\Gamma}(b,t),  \notag \\
\frac{d}{d t}{\Gamma}(b,t)&=&-i[h(b),{\Gamma}(b,t)]-\kappa{\Gamma}
(b,t)+\kappa{\Gamma}(a,t).  \label{Telegraph}
\end{eqnarray}
The desired stochastic average is $\langle\Gamma(t)\rangle_{s}={\Gamma}(a,t)+
{\Gamma}(b,t)$. Before proceeding with the general solution, we find it
worthwhile to briefly comment on the fast and slow (quasi-static disorder)
limit.

In the fast noise limit, when $1/\kappa$ is the shortest time scale, an
adiabatic elimination of the fast dynamics in lowest order perturbation
theory reduces the above equations to a master equation in Lindblad form, 
\begin{equation}
\frac{d}{dt}{\langle\Gamma \rangle}_{s}=-i\left[ h_{+},\langle \Gamma
\rangle _{s}\right] -\frac{1}{2\kappa }\left[h_{-},\left[ h_{-},\langle
\Gamma \rangle_{s}\right] \right] ,  \label{Telefast}
\end{equation}
with $h_{\pm}=[h(a)\pm h(b)]/2$. The first term describes a coherent
evolution with the average Hamiltonian $h_{+}\equiv\langle h\rangle_{s}$,
while the second term is the noise-induced damping, which scales as $1/\kappa$. Suppression of the dissipation by fast noises is called motional
narrowing in an atomic physics context \cite{CohenTannoudji1977}, and
corresponds to a Zeno effect \cite{Facchi2008}. If the \emph{average}
Hamiltonian is in the topological phase and thus supports Majorana edge
modes, the initial Majorana correlation will show slow dephasing due to the
Zeno effect - irrelevant if $a$ or $b$ \emph{per se} lies in the topological
or non-topological regime. On the other hand, in the limit of slow noise,
i.e. when the jump rate $\kappa\rightarrow 0$, we can ignore $\mathcal{L}$
in Eqs. (\ref{Master_density}) and (\ref{Telegraph}), $\mathcal{L}\rightarrow 0$, during long times $t\ll\kappa^{-1}$, so that for such
times we simply have to perform the (quasi-)static average of the system
dynamics.

\begin{figure*}[tb]
\centering
\includegraphics[width= 1.0\textwidth]{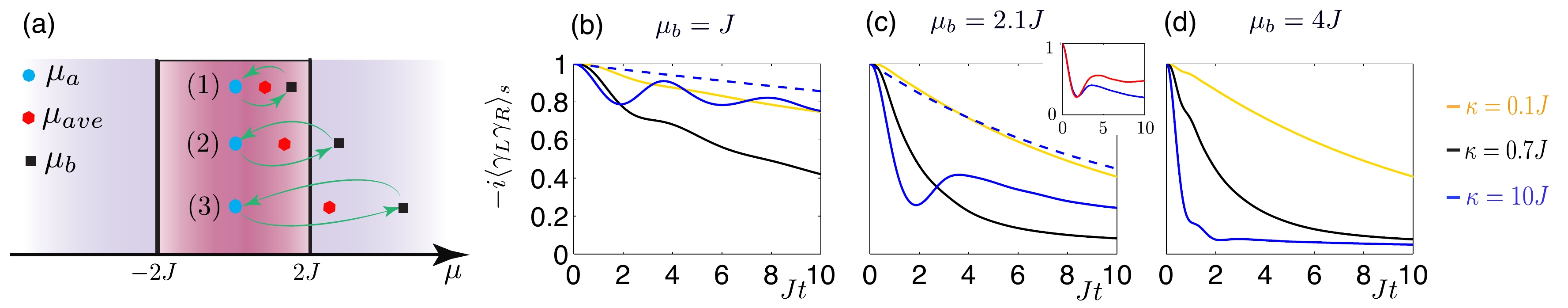}
\caption{Evolution of Majorana edge correlations in presence of globally
fluctuating chemical potential. (a) For a chemical
potential $\mu(t)$ flipping between $\mu_a$ and $\mu
_b$, there exist three jump scenarios (1)-(3) in the phase diagram
corresponding to the value of $\mu_b$. (b)-(d) Time evolution of
the Majorana edge correlation function $-i\langle\gamma_L\protect
\gamma_R\rangle_s$ for $\mu_a=0.2J$ and ({b}) $\mu_b=J$ (c) $\mu_b=2.1J$ and ({d}) $\mu_b=4J$. In each scenario, we
take three typical jump rates: $\protect\kappa=0.1J$, $\kappa=0.7J$,
and $\kappa=10J$, and consider a Kitaev wire of $N=60$ sites with
the pairing parameter $\Delta=0.8J$. We have considered two types of initial
conditions, which correspond to the ground state of Hamiltonian $H_a$ (solid
curves) and of average Hamiltonian $H_{\text{ave}}$  in the case of fast noises [blue dashed curves in (b) and (c)]. In the inset of (c), the dynamics for the noise with $\kappa=10J$ (blue curve) for $\mu_b=2.1J$ is compared with the quench dynamics (red curve), for a
quench in the chemical potential from $\mu=0.2J$ to $\mu=\mu_{\textrm{ave}}$. }
\label{Fig:global}
\end{figure*}

Below we will solve the full marginal density equations (\ref{Telegraph})
for Majorana edge correlation $-i\langle \gamma _{L}\gamma _{R}\rangle
_{s}=-\sum_{il}f_{L,i}f_{R,l}\langle\hat{\Gamma}_{il}\rangle _{s}$, with $f_{L/R,l}$ the Majorana wavefunction of the initial Hamiltonian in the
Majorana basis. Our discussion will focus on the special case of local and
global telegraph fluctuations of the chemical potential. As mentioned above,
such telegraph noise can represent a local two-level fluctuator in a solid
state setup, and global laser frequency noise in an atomic realization of
the Kitaev wires. We would like to stress, however, that even though we
consider here the simplest jump-like noise process and only the chemical
potential as a fluctuating parameter, our conclusions remain valid for a
more general scenario with colored noise and with several fluctuating
parameters.

We conclude this section with the remark that the considered Markovian noise
models give rise to first order noise correlation functions, which are
exponentials, or superposition of exponentials. We note however that
non-Markovian noise models can often be represented as projections of higher
dimensional Markov processes, which can also be solved by our techniques.
Another point to mention is that for non-quadratic Hamiltonians in one
dimension, the generalized master equation (\ref{Master_density}) can be
solved using the density matrix renormalization method \cite{Daley2004,Schollwock2005}, and the developed techniques therefore are
interesting in a much broader context for many-body systems driven by
colored noise.


\section{Edge and bulk dynamics in presence of global noise}

\label{sec:global}

We begin with discussing the time evolution of the Majorana edge correlation 
$-i\langle\gamma_{L}\gamma _{R}\rangle_{s}$ under the global fluctuations in
the chemical potential $\mu (t)$, when it flips between two values $\mu_{a}$
and $\mu_{b}$ (the corresponding Hamiltonian are $H_{a}$ and $H_{b}$,
respectively) with the jump rate $\kappa$. The statistic property of $\mu(t)$
is described by the mean value $\mu_{\text{ave}}=(\mu _{a}+\mu_{b})/2$, the
variance $\sigma^2=(\mu_{a}-\mu_{b})^2/4$, and the correlation time $
\tau_{c}=1/2\kappa$. We denote by $H_{\text{ave}}=\langle H[\mu(t)]\rangle_s$
the average Hamiltonian over the noise realization, which in the considered
case is of the form of a noise-free Kitaev Hamiltonian with $\mu=\mu_{\text{ave}}$. For the initial condition, we take $\mu(0)=\mu_{a}$ with $|\mu_{a}|<2J$ such that Hamiltonian $H_{a}$ is in the topological phase, and
assume the system is in the ground state with the Majorana edge-mode
correlation $-i\langle\gamma_{L}\gamma_{R}\rangle=1$. For the value $
\mu_{b}$, we consider three possible scenarios (see Fig. \ref{Fig:global}(a)): (1) $|\mu_{b}|<2J$, when the Hamiltonian $H_{b}$ is in the topological
phase (fluctuations within the topological phase); (2) $|\mu_{b}|>2J$ but $|\mu_{\text{ave}}|<2J$, when $H_{b}$ is in the non-topological phase but the
average Hamiltonian $H_{\text{ave}}$ remains in the topological phase
(fluctuations between topological and non-topological phases but on average
staying in the topological one); and (3) $|\mu_{b}|>2J$ and $|\mu_{\text{ave}
}|>2J$, when both Hamiltonians $H_{b}$ and $H_{\text{ave}}$ are
non-topological (large-amplitude fluctuation when staying on average in the
non-topological phase). The evolution of $-i\langle\gamma_{L}\gamma_{R}\rangle_{s}$ calculated on the basis of Eq. (\ref{Telegraph}) for these
three scenarios, is shown in Figs. \ref{Fig:global}(b) - \ref{Fig:global}(d), respectively, in the regimes of fast, intermediate, and slow jump rate $\kappa$.

We see that noise always leads to decay of Majorana correlations, but the
decay dynamics significantly depends on the amplitude and rate of the noise.
Non-surprisingly, we find the slowest decay of the Majorana correlations in
the scenario (1) when the fluctuating Hamiltonian always remains in the
topological phase. Strikingly, the dynamics in the scenario (2) shows the
same features, even though here we have jumps between topological and
non-topological phases: In both scenarios, we observe the fastest decay in
the regime with an intermediate jump rate ($\kappa=0.7J$, black curves in
Figs.~\ref{Fig:global}(b) and \ref{Fig:global}(c)), when $\kappa$ is
comparable with the energy gap and the band width ($\sim J\sim\Delta$) of
the Hamiltonian $H_{a}$; whereas, in both slow ($\kappa=0.1J$, yellow
curves) and the fast noise ($\kappa=10J$, blue curves) regimes, the decay is
much slower, and the system exhibits substantial Majorana correlations for
much larger times ($>10J^{-1}$). In contrast to this non-monotonic
dependence on the noise rate, the decay rate in the scenario (3) grows with $
\kappa$ (see Fig. \ref{Fig:global}(d)): for both intermediate and fast
fluctuations there are no visible Majorana correlations for times $
t\gtrsim10J^{-1}$, although for slow noise they survive for a much longer
time.

We now detail the analysis on the dynamical behavior of $-i\langle\gamma_{L}
\gamma _{R}\rangle_{s}$ in scenarios (1)-(3) in the slow, fast and
intermediate regimes of the noise, respectively. We start with slow
fluctuations when $\kappa\ll J\sim\Delta$, $\mu_{a(b)}$. In this case, the
quenches between $H_{a}$ and $H_{b}$ -- occuring at random instants -- on
average take place after a typical time $\kappa^{-1}$, which is much larger
than all characteristic time scales of the system. As is explained in App. 
\ref{App:quench}, on the time scale of several inverse band-widths of the
system ($\sim J^{-1}$) after a quench, the Majorana correlations relax to
asymptotic values which are determined by the overlap of the Majorana
edge-mode wave-functions for the Hamiltonians $H_{a}$ and $H_{b}$. Such
asymptotic values then remain constant till the next quench occurs. For the
scenarios (2) and (3), when $H_{b}$ is non-topological and has no edge
modes, the overlap is zero, and already the first quench completely destroys
the correlations. We therefore have $-i\langle\gamma_{L}\gamma_{R}\rangle_{s}\sim\exp(-\kappa t)$ for these scenarios (yellow curves in Figs.
\ref{Fig:global}(c) and \ref{Fig:global}(d)). On the other hand, for the
scenario (1) when $H_{b}$ has the zero mode, the overlap is non-zero. In
this case, each quench reduces the correlations by a factor of $G_{\infty}<1$
related to the overlap [see Eqs. (\ref{Glongtime}) and (\ref{Gf}) in App. \ref{App:quench}], resulting in a slower decay of the correlation $-i\langle\gamma_{L}\gamma_{R}\rangle_{s}\sim\exp[-\kappa(1-G_{\infty})t]$
(yellow curve in Fig. \ref{Fig:global}(b)).

\begin{figure*}[tbh]
\includegraphics[width=0.32\textwidth]{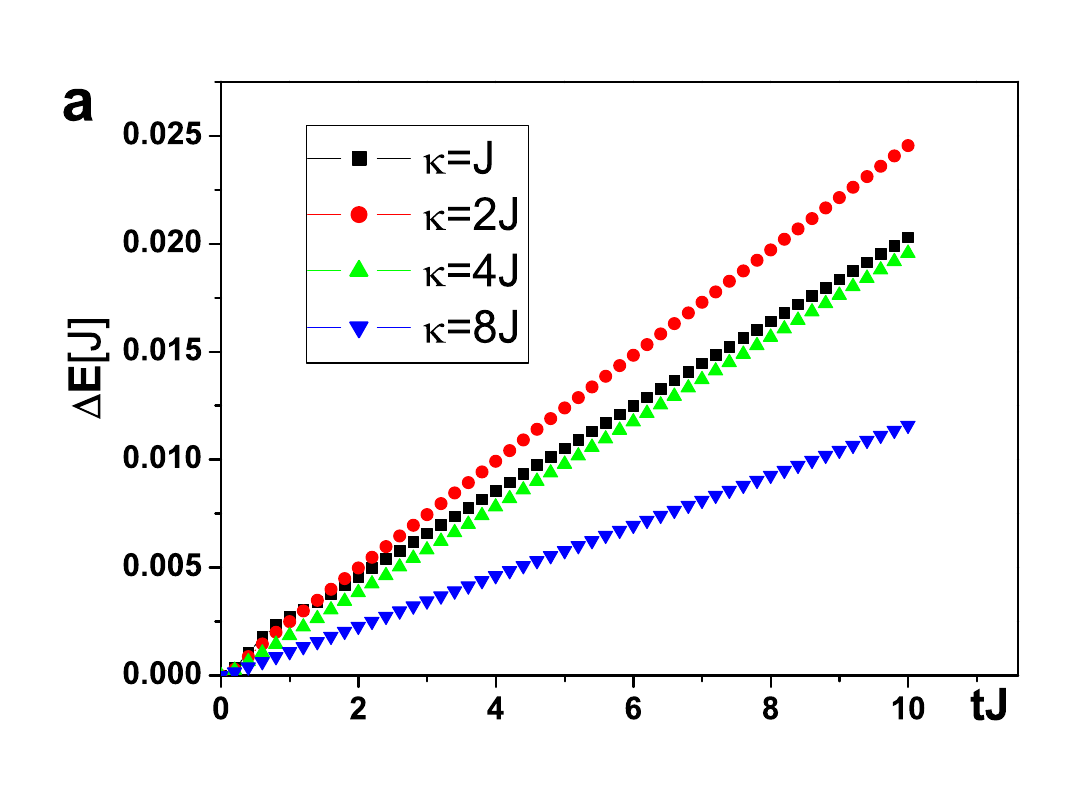}
\includegraphics[width=0.32\textwidth]{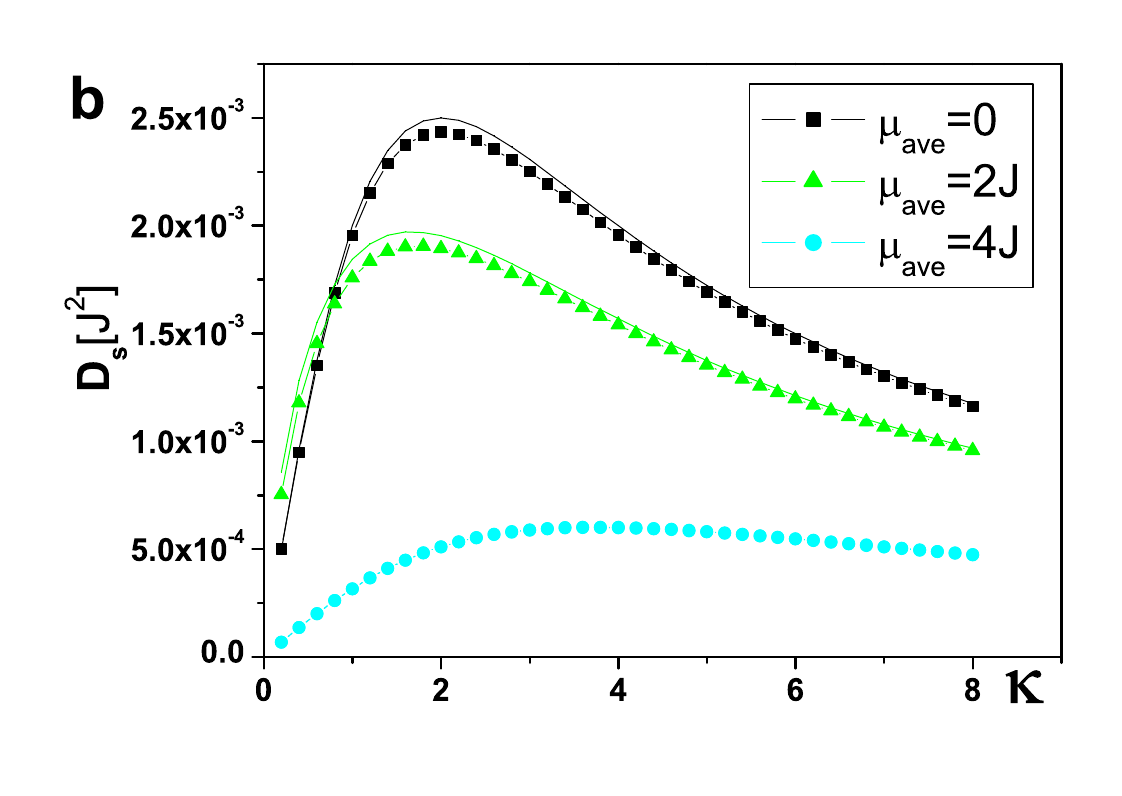}
\caption{Heating dynamics in the bulk. ({a}) The absorbed energy $\Delta E(t)$ in unit of $J$ as a function of time for $\mu _{\text{ave}}=0$ and
different $\kappa$; ({b}) Heating rate $D_{s}$ in unit of $J^{2}
$ as a function of $\kappa$ for different $\mu _{\text{ave}}$. Numerical results (dotted curves) are compared to the predictions from Eq. (\ref{Energy-growth}) (solid curves). For ({a})-({b}), we have fixed the noise variance $\sigma=|\mu _{b}-\mu _{a}|/2=0.1J$, and have chosen $\Delta=J$.}
\label{fig:bulk}
\end{figure*}

In the opposite regime of fast fluctuations ($\kappa\gg|\mu_{b}-\mu_{a}|$, $
J\sim\Delta$), the dynamic behavior is remarkably related to the Zeno effect
and to the quench problem. In this case, the evolution of $-i\langle\gamma_{L}\gamma_{R}\rangle_{s}$ can be explained based on Eq. (\ref{Telefast}) for the correlation matrix, which in the considered case
takes the form 
\begin{equation}
\frac{d }{d t}\langle\Gamma\rangle_{s}=-i[h_{+},\langle\Gamma\rangle_{s}]- 
\frac{\sigma^{2}}{2\kappa }[{N},[{N},\langle\Gamma\rangle_{s}]].
\label{Fastglobal}
\end{equation}
Here $\sigma=|\mu_b-\mu_a|/2$ as mentioned earlier, and the matrices $h_+$
and $N$ correspond to the Hamiltonian $H_{\text{ave}}$ and the total
particle number operator in the Majorana basis. Following from Eq. (\ref
{Fastglobal}), we see that:

(i) In the limit $\kappa\rightarrow\infty$, when the second {\textquoteleft
decay\textquoteright}-term can be neglected, Eq. (\ref{Fastglobal})
describes the dynamics of correlations after the quench from the initial
Hamiltonian $H_{a}$ to the \textit{averaged} one $H_{\text{ave}}$. As a
result, the asymptotic ($t\rightarrow\infty$) value of the Majorana
correlation function $-i\langle\gamma _{L}\gamma _{R}\rangle_{s}$ is
determined again by the overlap of the wave functions of the Majorana edge
modes (see App. \ref{App:quench}), but now for the Hamiltonians $H_{a}$ and $
H_{\text{ave}}$. For Hamiltonian $H_{\text{ave}}$ in the non-topological
phase [scenario (3)], there are no such modes, and after the quench $
-i\langle\gamma_{L}\gamma _{R}\rangle_{s}$ decays to zero on the time
scale of the order of the inverse band-width of $H_{\text{ave}}$. For $H_{\text{ave}}$ in the topological phase, [scenario (1) with topological $H_{b}$
and scenario (2) with \textit{non-topological} $H_{b}$], this mode exists
giving rise to a non-zero overlap and to a finite asymptotic value of the
correlations after the quench.

(ii) For a large but finite $\kappa$, the second term in Eq. (\ref{Fastglobal}) adds a slow decay on top of the quench dynamics, providing the
asymptotic behavior of $-i\langle\gamma_{L}\gamma_{R}\rangle_{s}$ shown in
Figs. \ref{Fig:global}(b) and \ref{Fig:global}(c). The short-time ($
t\lesssim J^{-1}$) behavior of the correlations, seen in the form of damped
oscillations on these figures, is sensitive to the details of the band
structures of $H_{a}$ and $H_{\text{ave}}$. In general, the {\textquoteleft
closer\textquoteright} these Hamiltonians are, the more pronounced are the
oscillations, and the less destructive effect has the quench on $-i\langle
\gamma_{L}\gamma_{R}\rangle_{s}$.

(iii) Note that Eq. (\ref{Fastglobal}) suggests the preparation of the
initial Majorana correlations with respect to the edge states of the
Hamiltonian $H_{\text{ave}}$, not $H_{a}$, which is practically also more
natural for fast fluctuations. In this case, the first term in Eq. (\ref
{Fastglobal}) has no destructive effects, and $-i\langle\gamma_{L}\gamma
_{R}\rangle_{s}$ shows slow decay due to the second term (dashed blue lines
in Figs. \ref{Fig:global}(b) and \ref{Fig:global}(c)). This slow decay of
Majorana correlations in presence of fast noises -- even for a sufficiently
large fluctuation amplitude outside the topological phase -- is a direct
consequence of the Zeno effect, which reduces the dynamics with fast
fluctuating parameters to a weakly damped dynamics with the averaged
Hamiltonian.

Finally, in the intermediate regime when the fluctuation rate is of the
order of typical energy scales in the system, $\kappa\sim J\sim\Delta$, $
\mu_{a(b)}$, one has optimal conditions for pumping excitations into the
system (heating), leading to the fastest decay of the Majorana correlations.
This can be seen by looking at the growth in the system energy under the
action of noises. To illustrate this heating dynamics, we consider the case
when $\mu(t)$ jumps between $\mu_{a}=\mu-\delta\mu $ and $\mu_{b}=\mu+\delta
\mu$ (such that $\mu_{\mathrm{ave}}=\mu$) with a rate $\kappa$, focusing on
the weak noise limit $\delta \mu\ll J$ and choosing $\Delta =J$. We assume
that initially the system is in the ground state of the \textit{average}
Hamiltonian $H_{\text{ave}}$ [but still $\mu (t=0)=\mu_{a}$], with energy $
E_{0}$ calculated with the corresponding initial density matrix $\rho
_{s}(0) $, and calculate the system energy gain $\Delta E(t)=\text{Tr}(\rho
_{s}(t)H_{\mathrm{ave}})-E_{0}$ for different $\mu$ and $\kappa$. A typical
time evolution of $\Delta E(t)$ for different noise jump rates is shown in
Fig.~\ref{fig:bulk}(a). There, the energy is seen to grow linearly, $\Delta
E(t)=D_{s}t$, after a short transition time ($t\lesssim J^{-1}$). Then after
a relatively long time (not shown) due to the small noise amplitude $
\delta\mu\ll J$, it saturates to the asymptotic value which depends on $\mu$
but not on $\kappa$ (infinite-temperature state). The heating rate $D_{s}$
depends on both $\kappa$ and $\mu$. Figure \ref{fig:bulk}(b) shows $D_s$ as
a function of the jump rate $\kappa$ for the values of chemical potential $
\mu=0$, $2J$, and $4J$, which correspond to the topological, critical, and
non-topological phases, respectively. We see that $D_{s}$ is a non-monotonic
function of $\kappa$, which is small when noise is slow or fast (Zeno
effect), and has a pronounced maximum for $\kappa\approx 2 - 4J$. The
maximum corresponds to the situation when the inverse noise correlation time 
$\tau_c^{-1}=2\kappa$, which determines the frequency width of the noise
correlation function, lies inside the band of bulk excitations (the exact
position depends on the band structure).

The linear growth of the energy for the considered times can be understood
by calculating the energy gain in the second-order perturbation
theory ($\hbar\equiv1$):
\begin{equation}
\Delta E(t)=\sum_{\nu}\varepsilon_{v}\left\vert M_{\nu}\right\vert^{2}\left\langle \left\vert \int_{0}^{t}d\tau\exp(i\varepsilon_{\nu}\tau
)\mu (\tau)\right\vert^{2}\right\rangle ,
\end{equation}
where $M_{\nu}$ is the matrix element of the number operator between the
ground state and the excited state $\left\vert \nu \right\rangle$ with the
energy $E_{0}+\varepsilon_{\nu}$. For the considered global perturbation, $\left\vert\nu\right\rangle$ simply corresponds to the state with two
single-particle excitations with momenta $k$ and $-k$, and the energy $
\varepsilon_{\nu }=2E_{p}=2\sqrt{(2J\cos pa+\mu )^{2}+4\Delta ^{2}\sin
^{2}pa}$ with $a$ being the lattice spacing. Performing the time derivative
and assuming the time $t$ being larger than the noise correlation time $\tau_{c}=(2\kappa )^{-1}$, we obtain the following expression for $D_{s}$:
\begin{equation}
D_{s}=\sum_{p}2E_{p}\left\vert M_{p}\right\vert ^{2}\int_{-\infty }^{\infty}d\tau \left\langle \mu (\tau )\mu (0)\right\rangle \exp (2iE_{p}\tau),\nonumber
\end{equation}
where the summation is over the Brillouin zone $p\in(-\pi/a,\pi/a]$. With the expression $\left\vert M_{p}\right\vert^{2}=\left\vert u_{p}v_{p}\right\vert^{2}=(\Delta^{2}\sin^{2}pa)/E_{p}^{2}$ for the
matrix element and $\left\langle \mu (\tau)\mu(0)\right\rangle=\mu^{2}+\sigma ^{2}\exp (-2\kappa \left\vert \tau \right\vert )$ for the noise
correlation function, we finally obtain ($k=pa$)
\begin{equation}
D_{s}=4\sigma^{2}\int_{-\pi }^{\pi }\frac{dk}{2\pi}\frac{\kappa }{\kappa^{2}+E^2_{k}}\frac{\Delta^{2}\sin^{2}k}{E_{k}},  \label{Energy-growth}
\end{equation}
for the energy-growth coefficient. The above expression is plotted as solid
lines in Fig. \ref{fig:bulk}(b) and is in very good agreement with numerical
data.

\begin{figure}[tb]
\centering
\includegraphics[width=1.0\columnwidth]{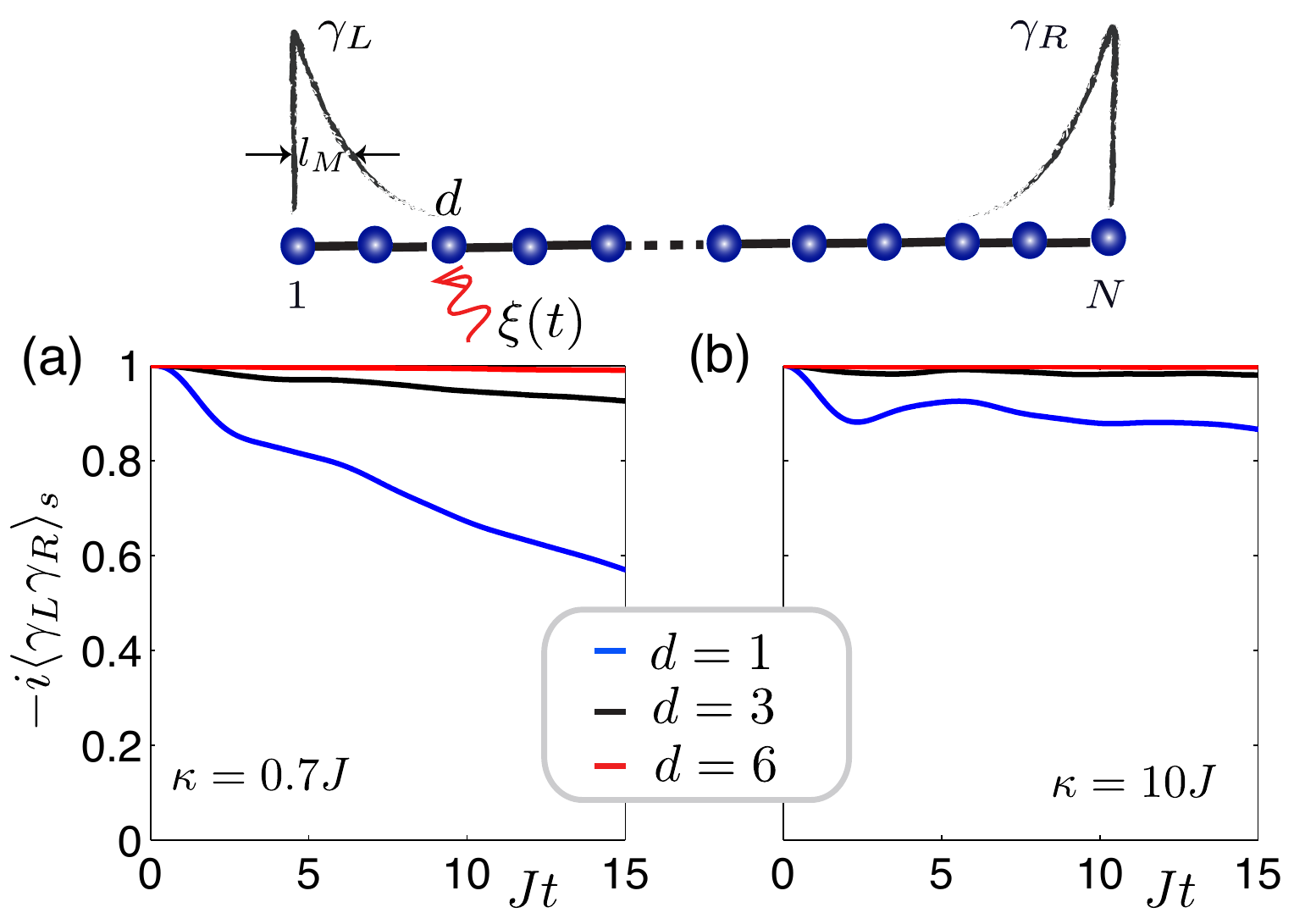}
\caption{The effect of local noise at different locations in a Kitaev wire on Majorana edge correlations. (a)-(b) Time evolution of $
-i\langle\gamma_L\gamma_R\rangle_s$ when a noise $\xi(t)$ arises in the local chemical potential at three different
sites ($d=1,3,6$), with a jump rate (a) $\kappa=0.7J$ and (b) $\kappa=10J$. We take $\xi_a=0$ and $\xi
_b=0.8J$, $N=60$, $\Delta=0.4J$, and $\mu=0.4J$. For this parameter
choice, the localization length of the Majorana modes is $l_M\approx 2.5a$, with $
a$ being the lattice constant. }
\label{Fig:local}
\end{figure}

\begin{figure}[tb]
\centering
\includegraphics[width=1.0\columnwidth]{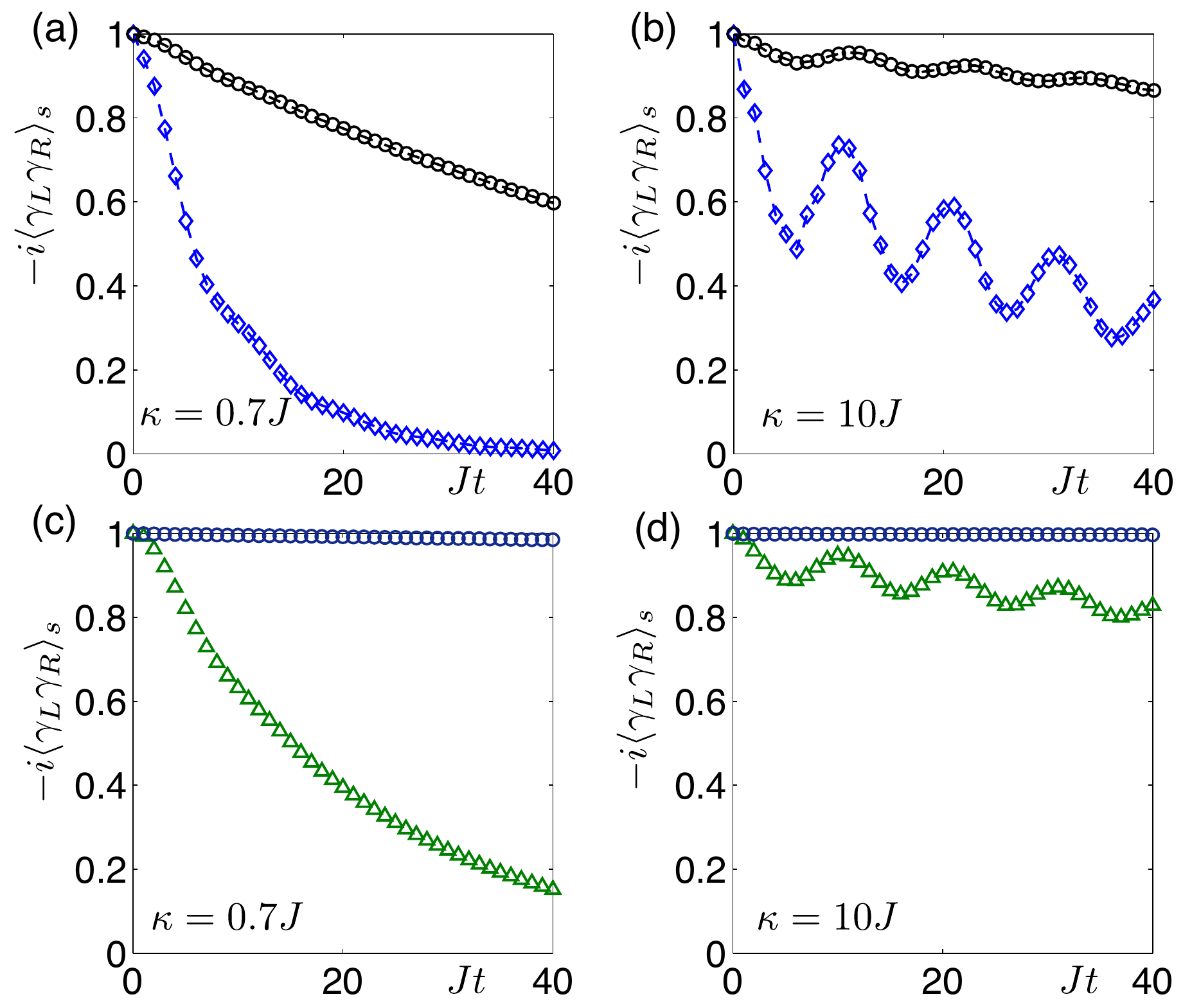}
\caption{(a)-(b) The effect of a large-amplitude noise $\xi(t)$ in the local chemical potential on the dynamics of Majorana edge correlation $-i\langle\gamma_L\gamma_R\rangle_s$, when the local noise $\xi(t)$ arises at site $d=3$ (blue curves)
and $d=6$ (black curves), flipping between $\xi_a=0$ and $\xi_b=4.4J$. (c)-(d) Evolution of the edge correlation $-i\langle\gamma_L\gamma_R\rangle_s$ when noise induces at random instants a splitting in the wire, between sites $3$ and $4$ (green curves) and between sites $6$ and $7$ (blue curves).  In (a)-(d), we take two different noise jump rates, $\kappa=0.7J$ (left panels)
and $\kappa=10J$ (right panels). Other parameters are the same as Fig. \ref{Fig:local}. }\label{Fig:split}
\end{figure}


\section{Local noise}

\label{sec:local}

We next discuss the effects of the local noise on the Majorana correlations $
-i\langle\gamma _{L}\gamma _{R}\rangle_{s}$, which we model by adding a
fluctuating part to the chemical potential on the site $d$, such that $\mu
_{j}(t)=\mu +\delta _{j,d}\xi (t)$ with the amplitude $\xi (t)$ being
described by the telegraph noise: $\xi (t)$ randomly flips between $\xi _{a}$
and $\xi _{b}>0$ with the rate $\kappa $, and $\xi (0)=\xi _{a}$. In this
case, an exponential localization of the Majorana modes near the edges with
the localization length $l_{M}$ leads to a very strong dependence of the
effects of the noise on $d$. This is because the decay of the Majorana
correlations is caused by noise-induced changes in the population of the
associated non-local fermionic zero-energy mode, with the corresponding
matrix element being proportional to the value of the edge-mode wave
function on the noisy site $d$. The results of our calculations for
intermediate and fast noises, Figs. \ref{Fig:local}(a) and \ref{Fig:local}
(b), respectively, clearly show this dependence: the decay is the fastest
for $d<l_{M}$, for $d\sim l_{M}$ it is already substantially less, and is
exponentially small for $d>l_{M}$.

Similar to the case of a global noise, the reduction of the decay in the
fast-noise regime ($\kappa\gg J$, $\Delta$, $\mu$) is due to the Zeno
effect, as follows from Eq. (\ref{Telefast}) which now takes the form 
\begin{equation}
\frac{d}{d t}\langle\Gamma\rangle_{s}=-i[h_{+},\langle\Gamma\rangle_{s}]- 
\frac{\sigma^2}{2\kappa }[n_{d},[n_{d},\langle\Gamma\rangle_{s}]],  \notag
\end{equation}
where $\sigma=|\xi_b-\xi_a|/2$, the matrix $n_d$ corresponds to the local density operator $a_d^\dag a_d$ at site $d$ in the Majorana basis, and the second
term adds a slow decay on top of the result of the quench described by
the first term. The average Hamiltonian in this case contains the \textit{static} impurity potential $V_{d}=(\xi _{a}+\xi _{b})/2$ on the site $d$,
which results in just modification of the Majorana edge modes, and hence in
a finite asymptotic value of $-i\langle\gamma_{L}\gamma_{R}\rangle_{s}$
after the quench. The small ($\sim\kappa^{-1}$) decay rate is extra reduced,
as compared to the global noise, for $d\gtrsim l_{M}$ due to the smallness
of the edge-mode wave functions on the site $d$. For $d>l_{M}$, this gives
an exponentially small decay rate such that the Majorana correlations are
practically immune to the noise (Fig. \ref{Fig:local}).

If we increase the amplitude of the noise to larger values, see
Figs. \ref{Fig:split}(a) and \ref{Fig:split}(b) for $\xi_{b}=4.4J$, the oscillating behavior for fast
noise become much more pronounced. In this case, the strong static impurity
potential in the average Hamiltonian splits the wire into two pieces which are
weakly coupled through the impurity site (Josephson junction). If the
coupling would be zero, one would have an extra pair of Majorana modes at
the edges adjacent to the impurity, and the corresponding fermionic
zero-mode. For a small but finite coupling, the energy of this mode is
finite, and the oscillations seen in this case correspond to the energy of
this mode. Similar oscillatory behavior is observed when the fast noise randomly splits the wire into two parts, for example, when the local hopping $J_d$ and the pairing $\Delta_d$ amplitudes (between site $d$ and $d+1$) jump simultaneously between the finite values and zero, as shown in Figs. \ref{Fig:split}(c) and \ref{Fig:split}(d). In this case of fast noise, the average Hamiltonian has a ``weak link", and the oscillation frequency seen in Fig. \ref{Fig:split}(d) corresponds to the energy of the fermionic mode localized at this link. Note that the amplitude of the oscillations is related to the overlap between the Majorana edge mode and the wave function of the low-energy fermionic mode localized on the ``defected" site or link, and rapidly decreases with increasing distance between the modes.

Note that for the ideal Kitaev chain ($J=\Delta$ and $\mu=0$), the two
Majorana modes $\gamma_{L}=c_{1}$ ($\gamma_{R}=c_{2N}$) locate on the
leftmost (rightmost) sites, such that $-i\langle\gamma _{L}\gamma_{R}\rangle_{s}=-(\Gamma_{s})_{1,2N}$. In this case, the dynamics of the
Majorana correlations is completely uncoupled from that of the bulk -- the
Majorana correlations in this ideal case are absolutely insensitive to what
happens in the bulk.


\section{Competition between noise and adiabaticity in Majorana Transport}

\label{sec:move}

\begin{figure}[tb]
\centering
\includegraphics[width=1.0\columnwidth]{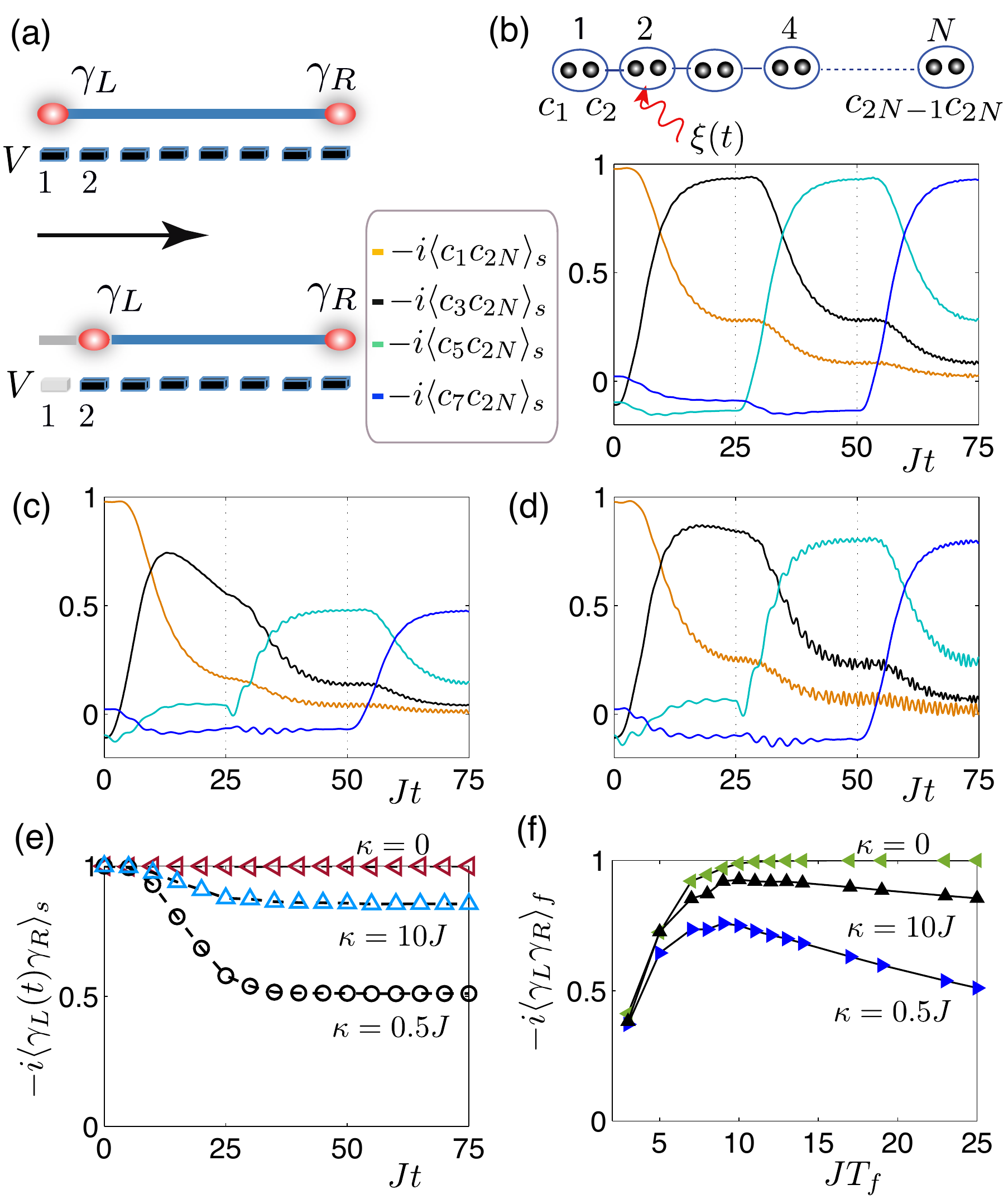}
\caption{Fate of a Majorana moving through a local noise. (a) A move of the Majorana edge mode $\gamma_L$
from site $1$ to site $2$ by adiabatically tuning the local potential $V$ in
a time interval $T_f$. (b)-(e) Time evolution of the correlation between
the edge Majorana operators $-i\langle c_{2j-1}c_{2N}\rangle_s$ for $
j=1,2,3,4$, when $\gamma_L$ is adiabatically transported from site $
1 $ to site $4$ through a local noise $\xi(t)$ at site $d=2 $ with (b) $\kappa=0$ ({c}) $\kappa=0.5J$ and ({d}) $\kappa=10J$. (e) The time evolution of the correlations between the Majorana edge
modes $-i\langle\gamma_L(t)\gamma_R\rangle_s$ for different 
$\kappa$. ($JT_f=25$ for ({b})-({e})). ({f}) The remaining Majorana
correlations $-i\langle\gamma_L\gamma_R\rangle_f$ as a
function of $T_f$ for different $\kappa$, after $\gamma_L$ is
moved to site $4$ (in total times of $3T_f$). For ({b})-({f}), we have chosen $\xi_a=0$, $\xi _b=0.8J$, $N=40$, $\Delta=0.8J$, and $\mu=0.2J$.}
\label{Fig:move}
\end{figure}

\begin{figure*}[tb]
\centering
\includegraphics[width= 1.0\textwidth]{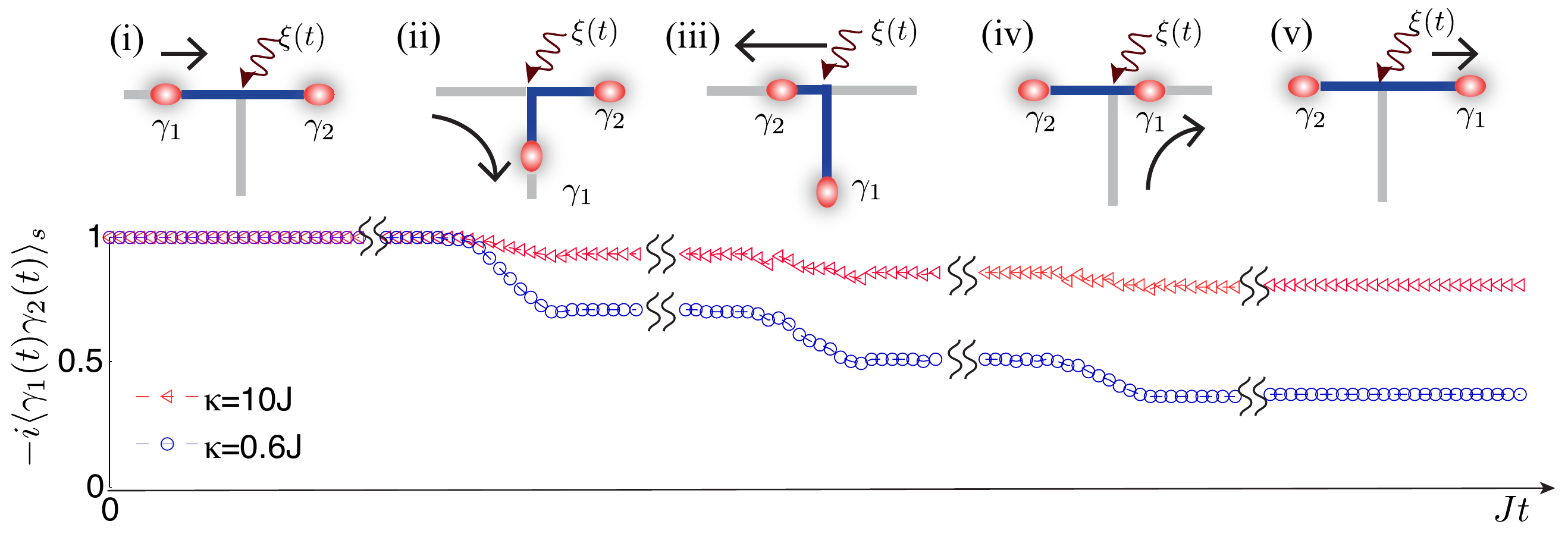}
\caption{Majorana braiding on a noisy T-junction, where a local noise $\xi(t)$ occurs
in the common joint connecting three wire segments. Upper
panel: schematics of an exchange of two Majoranas $\gamma_1$ and $\gamma _2$ following Ref. \cite{Alicea2011}: (i) Initially
the horizontal wire is topological supporting two Majorana edge modes $\gamma_1$ and $\gamma_2$ at the ends, while the vertical
wire is non-topological. (ii) $\gamma_1$ is moved through the
junction to the bottom of the vertical wire, while $\gamma_2$ is
fixed at the right end of the horizontal wire. (iii) $\gamma_2$ is
moved all the way to the left end of the horizontal wire, while $\gamma_1$ is fixed. (iv) $\gamma_1$ is moved upward through the
junction and then rightward. (v) At the end of the exchange, the wire
returns to its original configuration, with $\gamma_1\rightarrow\gamma_2$ and $\gamma_2\rightarrow-\gamma_1$. Lower
panel: Evolution of the Majorana edge correlations $-i\langle\gamma
_1(t)\gamma_2(t)\rangle_s$ in each step of the exchange processes,
when a local noise $\xi(t)$ in the junction jumps between $0$ and $0.7J$
at a jump rate $\kappa=0.6J$ and $\kappa=10J$, respectively.
For other parameters, the pairing parameter of the horizontal and vertical
wires are chosen as $\Delta_x=i\Delta_y=0.7J$.  We take $\mu=0.1J$ in the topological segment and $\mu=-4J$ in the nontopological segment of the wires,  and $JT_f=18$.}
\label{Fig:T-junction}
\end{figure*}

Equipped with above understanding of the non-equilibrium dynamics of a noisy
Kitaev wire, we now discuss the effect of a local noise on the Majorana edge
correlations during the adiabatic transport \cite
{Alicea2011,Halperin2012,Romito2012} -- an essential building block for the
braiding operations. As we will see, in the presence of a noise, the
adiabaticity of the transport -- required for preserving the information
encoded in the Majorana correlations -- confronts with the finite life-time
of the correlations, and the competition of these two factors establishes an
optimal operation time.

Following Ref. \cite{Alicea2011}, we will move the left Majorana edge mode
by ``pushing" it to the right via
adiabatically switching on local potentials on the corresponding sites. For
example (see Fig. \ref{Fig:move}(a)), the move of $\gamma _{L}$ from site $1$
to site $2$ can be achieved by applying the local potential $\lambda (t)V$
at site $1$ [an extra term $\lambda (t)Va_{1}^{\dag}a_{1}$ in the
Hamiltonian], where $V\gg2J$ and $\lambda (t)$ increases monotonically from 
$\lambda (t<0)=0$ to $\lambda (t>T_{f})=1$ during the time interval $
[0,T_{f}]$ with $T_{f}$ being much larger than the inverse energy gap, $
T_{f}\gg J^{-1}$, $\Delta^{-1}$, $\mu^{-1}$. (In our calculations we use $
\lambda (t)=\sin ^{2}[(\pi /2T_{f})t]$.) Further moves can be achieved by
applying the same protocol successively to sites $2$, $3$, \ldots. In Fig. \ref{Fig:move}(b) we show the evolution of the correlations $-i\langle c_{2j-1}c_{2N}\rangle$ with $j=1,2,3,4$, during the adiabatic move of $
\gamma_{L}$ from site $1$ to site $4$ for $T_{f}=25J^{-1}$ (in total time $3T_{f}$) and in the absence of the noise. (The correlation $-i\langle\gamma_{L}(t)\gamma_{R}\rangle$ between the actual edge modes remains unchanged
from its initial value $1$, see also the curve with $\kappa =0$ in Fig. \ref
{Fig:move}(e)).

The same correlations in the presence of the local noise $V_{d}[\xi(t)]=\xi(t)n_{d}$ at site $d$ [here $\xi(t)$ flips again between $\xi _{a}$ and $\xi_{b}$ at a rate $\kappa$] are shown in Figs. \ref{Fig:move}(c) ($\kappa
=0.5J$) and \ref{Fig:move}(d) ($\kappa =10J$) for $d=2$. The corresponding
behavior of the correlation $-i\langle\gamma_{L}(t)\gamma _{R}\rangle_{s}$
between the edge modes [$\gamma_{L}(t)$ is moving and $\gamma _{R}$ is
fixed] is presented in Fig. \ref{Fig:move}(e). We clearly see deviations
from the noise-free case, which are significant for $\kappa=0.5J$ and small
for $\kappa =10J$ (Zeno effect), but these deviations take place mostly when
the Majorana mode moves to the noisy site (for $t$ from $0$ till $T_{f}$ in
the considered example). After that, the Majorana correlations do not
exhibit any visible decay and repeat the pattern of the noise-free case but
with the reduced amplitudes. This behavior follows from the localized
character of wave functions of the Majorana edge states: the correlations
between them are influenced by the local noise only when the moving Majorana
mode and the noisy site are within the localization length $l_{M}$. (Note
that the extend of the edge-mode wave function in the {\textquoteleft{non-topological}\textquoteright} part of the wire -- the sites with non-zero
local potential $V$-- is also non-zero but very small for $V\gg 2J$. As a
result, a noisy site in this part of the wire has no effect of the
correlations.)

Above results also imply that, in order to minimize the destructive effect
of the noise on Majorana edge correlations, the move through the noisy site
has to be performed with the fastest speed -- the requirement which is
opposite to the adiabaticity condition for the transport. As a result, there
exists an \textit{optimum speed} of transport (optimum $T_{f}$) for each $
\kappa$. This is illustrated in Fig. \ref{Fig:move}(f) which shows the
remaining correlations (after the total move) as a function of $T_{f}$ for
different $\kappa$: the decrease in the correlations for small $T_{f}$ is
due to non-adiabatic effects, while for large $T_{f}$ it is due to
accumulated action of the noise. The proper choice of $T_{f}$ can
substantially reduces the loss of correlations, especially in the
intermediate-noise regime.

Similar consideration is also applicable to the global noise. However, in
this case the destructive effect of the noise is independent of the position
of the Majorana modes, so that an entire time of the operation should be
within the life-time of correlations, see Fig. \ref{Fig:global}.


\section{Fidelity of braiding in a noisy wire network}

\label{sec:braid}

\begin{figure}[tb]
\centering
\includegraphics[width=1.0\columnwidth]{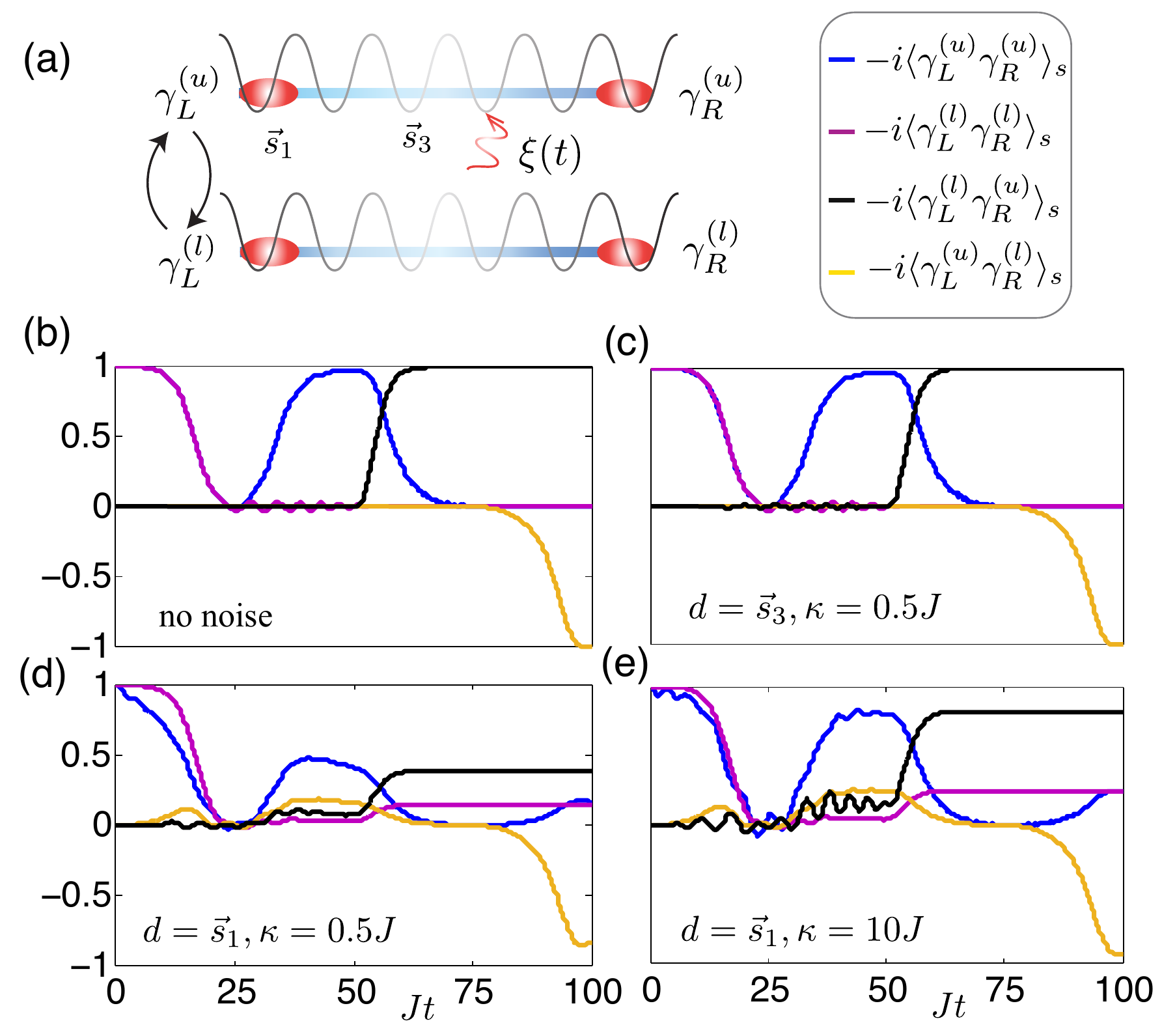}\newline
\caption{Braiding dynamics of Majoranas in a noisy atomic wire network. (a) Cold-atom implementation of braiding based on Refs. \cite
{Kraus2013,Laflamme2014}, where the Majorana modes $\protect\gamma_L^{(u)}$
and $\gamma_L^{(l)}$ on the left ends of the two wires are
exchanged. ({b})-({e}) Time evolution of the Majorana correlation functions
for ({b}) noise-free case ({c}) a local noise at site $\vec{s}_3$ of the upper wire with $\kappa=0.5J$, and ({d})-({e}) a local noise at site $\vec{s}_1$ of the upper wire with
(d) $\kappa=0.5J$ and ({e}) $\protect\kappa=10J$. For ({c})-({e}),
we choose $\xi_a=0$ and $\xi_b=0.6J$. For other parameters, $\Delta^{(u)}=\Delta^{(l)}=0.7J$ and $\mu^{(u)}=\mu^{(l)}=0$, 
$N^{(u)}=N^{(l)}=32$. Following Refs. \cite{Kraus2013,Laflamme2014},
the braiding is realized in four steps, where we choose the operation time $t_f$ for each step as $Jt_f=25$ for ({b})-({e})).}
\label{Fig:braid_coldatom}
\end{figure}

Finally, we study the effects of the noise on the Majorana braiding
(exchange) -- the operation which for the two modes $\gamma_{1}$ and $\gamma _{2}$ corresponds (up to a phase) to the unitary operator $U=\exp{[-(\pi /4)\gamma_{1}\gamma_{2}]}$ and results in the transformation $\gamma_{1}\rightarrow\gamma_{2}$, $\gamma_{2}\rightarrow-\gamma_{1}$, showing
non-Abelian character of Majorana fermions \cite{Alicea2011,Ivanov2001}. We
consider two proposed braiding scenarios: (i) in the T-junction for the
solid-state heterostructures, see Ref. \cite{Alicea2011}, and (ii) in the
wire-networks for cold-atom systems, see Refs. \cite{Kraus2013,Laflamme2014}.

We first consider the T-junction with two Majorana edge modes $\gamma_{1}=\gamma _{L}$ and $\gamma_{2}=\gamma_{R}$ which we braid by moving
them, see Fig. \ref{Fig:T-junction}, in accordance with the protocol from
Ref. \cite{Alicea2011}. We choose initially $-i\langle\gamma_{1}\gamma_{2}\rangle=1$ and follow the evolution of this correlation during the
protocol. Without noise, it remains unchanged during the entire braiding,
provided we move the modes, say $\gamma_1$, adiabatically (we choose $T_{f}=18J^{-1}$). In the presence of local noise, from the previous results
we expect the decrease of $-i\langle\gamma_{1}(t)\gamma _{2}(t)\rangle_{s}$
each time when the Majorana mode passes the noise site. This is demonstrated
in Fig. \ref{Fig:T-junction} for the case when the noise-source is located
at the common point of the three legs forming the T-junction: here the
Majorana modes have to cross the noisy site three times, and each crossing
results in the decrease of $-i\langle\gamma _{1}(t)\gamma _{2}(t)\rangle_{s}$. For the noisy site located in one of the legs, only two crossings occur
with the two corresponding drops in $-i\langle\gamma_{1}(t)\gamma_{2}(t)\rangle_{s}$, resulting in a higher fidelity of the braiding
operation.

For the braiding in an atomic wire network, we consider two wires (see Fig. \ref{Fig:braid_coldatom}(a)): the upper one (u) and the lower one (l), each
having a pair of Majorana modes ($\gamma _{L}^{(u)},\gamma _{R}^{(u)}$) and ($\gamma _{L}^{(l)},\gamma _{R}^{(l)}$). The braiding protocol from Refs. \cite{Kraus2013,Laflamme2014} involves operations only on one side (say,
left) of the network, and the mode to be braided, $\gamma_{1}=\gamma
_{L}^{(l)}$ and $\gamma _{2}=\gamma _{L}^{(u)}$, are also located on the
same side. As a result, the protocol will be only sensitive to noise located
close to the left side of the network. Fig. \ref{Fig:braid_coldatom}(b)
shows the evolution of correlations between various Majorana modes during
braiding in the absence of the noise. The evolution of the same correlations
with the telegraphic-noise source with $\kappa =0.5J$  on the third site ($\vec{s}_{3}$) and on the first site ($\vec{s}_{1}$) of the upper wire are
presented in Figs. \ref{Fig:braid_coldatom}(c) and \ref{Fig:braid_coldatom}
(d), respectively. They clearly show the above mentioned feature of the
protocol. Notably, for a fast noise, even when the noise source is located
on the first site $\vec{s}_{1} $, one has much less noise-induced
decoherence whence higher fidelity, see Fig. \ref{Fig:braid_coldatom}(e).

\section{Conclusions and Outlook\label{sec:conc}}

To summarize, we have studied the decoherence of Majorana edge correlations
and braiding dynamics in colored Markovian noises preserving parity
symmetry. Our analysis relies on a technique for solving quantum many-body
dynamics when the system parameters undergo local or global fluctuations
modelled by classical stochastic processes with arbitrary correlation time.
Our studies on noisy Kitaev wires show that, while the noise always gives
rise to the decay of the correlations between Majorana edge-states, there
are several parameter regimes where the life-time of the correlations
remains sufficient for quantum manipulations with Majorana fermions, even
without error-corrections. This includes the cases of slow global
noise and generic local noise in the bulk, and in particular, the case of fast noise where decoherence can be suppressed due to motional narrowing, also known as the Zeno effect. These
results further allow us to optimize the manipulation protocols of Majoranas
in both the solid-state and cold-atom settings. Our presentation is for
two-level telegraph noises in chemical potentials, but the essential
features of noise dynamics are also seen in the colored Gaussian noises (the
lattice model), and in other types of noises, e.g. the phase fluctuation in the pairing parameters.

The present study and the development of techniques to treat the effect of
noise from static disorder to the rapid fluctuation limit, should also be
seen in the broader context of dynamics of correlations in an interacting
many-body quantum system in the presence of random fluctuations. The effect
of random fluctuations, either as spatial disorder or temporal noise, on the
properties of quantum many-body systems is a long-standing and important
problem. Spatial static disorder underlies phenomena like Anderson
localization, and quantum many-body localization-delocalization transition
in the presence of interaction, with typically short-range correlations in
the localized phase. The time-dependent random fluctuations introduce
temporal decoherence and possible heating, resulting in finite temporal
correlations. Whether the combination of random fluctuations and
interparticle interactions could lead to some interesting long-range
dynamics in the space-time domain, is an open and intriguing question. The
considered model, being formally quadratic, implicitly contains the effects
of interparticle interactions in the form of a paring term which is
responsible for the existence of the non-trivial topological phase with
non-Abelian Majorana states. Due to topologically protection, these states
and the long-range correlations between them survive the static disorder,
and, therefore, the considered model provides a simple and tractable example
from a very special class of topological system, both interacting and
non-interacting, with correlations robust against static disorder. The
results of our paper provides therefore a possible scenario for behavior of
such systems in the presence of a temporal noise.

\begin{acknowledgments}
We acknowledge helpful discussions with A. J. Daley, H. Pichler, J. Budich, M. Heyl, P. Hauke, and T. Ramos. This project was supported by the ERC Synergy Grant UQUAM and the SFB FoQuS (FWF Project No. F4016-N23). Y. H. acknowledges the support from the Institut f\"{u}r Quanteninformation GmbH. 
\end{acknowledgments}

\bigskip

\appendix

\section{Derivation of the generalized master equation}\label{App:derive}

Following Ref. \cite{Peter1981}, here we derive the generalized master equation (\ref{Master_density}) for the marginal density matrix $\rho(X,t)$ in the main text. Denoting $
A[X(t)]\rho(t)\equiv-i\Big[H[X(t)], \rho(t)\Big]$, for each noise
realization we can solve the multiplicative stochastic equation $\dot{\rho}(t)=A[X(t)]\rho(t)$ for the density matrix $\rho(t)$ given an initial one $\rho(t_0)$ at $t_0$ with $X(t_0)=X_0$. The formal solution can be written as
$\rho(t)=\sum_{n=0}^\infty\rho_n(t)$, with $\rho_n(t)=\int_{t_0}^tdt_n...\int_{t_0}^{t_2}dt_1\prod_{i=1}^{n} A[X(t_i)]\rho(t_0)$ for $t\ge t_n\ge...\ge t_0$ and $\rho_0(t)=\rho(t_0)$. Thus from Eq. (\ref{Marginal}) we have $\rho(X,t)=\sum_{n=0}^\infty\rho_n(X,t)$ with $\rho_n(X,t)=\langle\rho_n(t)\delta(X-X(t))\rangle_s$, where the stochastic average can be straightforwardly performed using joint probability densities $P(X,t;X_n,t_n;...;X_0,t_0)$. From the defining property of a Markov process, that is the factorization property of the conditional probability densities \cite{Gardiner2010Stochastic}, we can write\begin{widetext}
\begin{equation}
\rho_n(X,t)=\int_{t_0}^t dt_n..\int_{t_0}^{t_2}dt_1 \int dX_{n}..\int dX_0A(X_n)...A(X_1)\rho(t_0)P(X, t|X_n,t_n)P(X_n,t_n;..;X_0,t_0). \label{Fnt}
\end{equation}
\end{widetext}
Here $P(X,t|X_n,t_n)$ is the conditional probability for finding $X(t)=X$
given $X(t_n)=X_n$ at the earlier time $t_n<t$. The significance of
Eq. (\ref{Fnt}) is that the time dependence in the stochastic
parameter $X(t)$ is now transferred into $P(X,t|X_n,t_n)$. Then by using the Chapman-Kolmogorov equation (\ref{CK})
for the evolution of $P(X, t|X^{\prime},t^{\prime })$ ($t^{\prime
}<t$), together with $P(X,t|X^{\prime },t)=\delta(X-X^{\prime })
$, we obtain 
\begin{eqnarray}
\dot{\rho}_n(X,t)=A(X)\rho_{n-1}(X,t)+\mathcal{L}(X)\rho_n(X,t),
\label{Fnt1}
\end{eqnarray}
and $\dot{\rho}(X,t)=\sum_{n=0}^\infty \dot{\rho}_n(X,t)$ readily gives Eq. (\ref{Master_density}).

\section{Fast fluctuation and
quasi-static limits}

\label{App:limit}

Below we solve the generalized master equation (\ref{Master_density}) for the average density matrix $\rho_s(t)=\langle\rho(t)\rangle_s$ in two limiting cases of a stationary colored Markovian noise: (1) the fast fluctuation limit and (2) the quasi-static limit. (As above we will use the notation $A(X)\rho(X,t)=-i[H(X), \rho(X,t)]$.)

\textit{Fast fluctuation limit}-- In this case, a master equation for $\rho_s(t)$ can be derived using the eigenfunction expansion method \cite{Gardiner2010Stochastic,Peter1981}. Denoting the left (right)
eigenfunctions of the noise operator $\mathcal{L}$ as $P_\lambda(X)$ [$Q_\lambda(X)$], with $\mathcal{L}P_\lambda(X)=-\lambda P_\lambda(X)$ [$\mathcal{L}^\dag Q_\lambda (X)=-\lambda Q_\lambda (X)$], we expand
\begin{equation}
\rho(X,t)=\sum_{\lambda}P_{\lambda}(X) C_{\lambda}(t). \label{Expansion}
\end{equation}
Note that $P_0(X)=P_s(X)$ represents a stationary distribution with $\mathcal{L}P_s(X)=0$,
and $Q_0(X)=1$. Denoting $(Q_\lambda, AP_{\lambda^{\prime }})=\int dX
Q_{\lambda}(X)A(X)P_{\lambda^{\prime }}(X)$ and using the orthogonal condition $(Q_{\lambda}, P_{\lambda^{\prime
}})=\delta_{\lambda,\lambda^{\prime }}$, the expansion coefficient $C_\lambda(t)$ in Eq.~(\ref{Expansion}) is derived as $
C_\lambda=\int dX Q_\lambda(X)\rho(X,t)$. Importantly, we identify
\begin{equation}
\rho_s(t)=C_0(t)\equiv\int dX\rho(X,t),  \label{Rho0}
\end{equation}
which is just the desired average density matrix. 

We thus want to derive $\dot{\rho}_{s}=\dot{C}_{0}(t)$. Substituting Eq. (\ref{Expansion}) into the generalized master equation (\ref{Master_density}), we find
\begin{eqnarray}
\!\!&&\dot{\rho}_{s}=A_{\text{ave}}\rho _{s}+\sum_{\lambda \neq
0}(Q_{0},AP_{\lambda })C_{\lambda}, \label{Rho0t}\\
\!\!&&\dot{C}_{\lambda}=-\lambda
C_{\lambda}+(Q_{\lambda },AP_{0})\rho _{s}+\sum_{\lambda^{\prime}\neq0}(Q_{\lambda},AP_{\lambda
^{\prime}})C_{\lambda^{\prime}}.  \label{Rhont}
\end{eqnarray}
with $A_{\text{ave}}=\int dXA(X)P_{s}(X)$. For fast fluctuations when the damping rate $\sim\lambda$ is large, we can eliminate the fast dynamics of $C_{\lambda }(t)$
[see Eq. (\ref{Rhont})] on a time scale $t\gg\lambda^{-1}$ using the technique of adiabatic eliminations \cite{Gardiner2010Stochastic}. In doing so, Equation (\ref{Rho0t}) becomes (in linear order of $1/\lambda$) 
\begin{equation}
\dot{\rho}_{s}(t)=\left( A_{\text{ave}}+\mathcal{D}\right) \rho _{s}(t).\label{Rho0t1}
\end{equation}
Here the operator $\mathcal{D}$ is defined by $\mathcal{D}\rho_s\equiv\int_0^td\tau\langle A(\tau),A(0)\rangle_s\rho _{s}$,  with $\langle A(\tau), A(0)\rangle_s=\sum_{\lambda\neq 0}[\int dX A(X)P_\lambda(X)]^2e^{-\lambda\tau}$ the stationary variance \cite{Gardiner2010Stochastic} and $t\gg\lambda^{-1}$.  Equation (\ref{Rho0t1}) is the familiar master equation: the first term corresponds to a coherent evolution with the average Hamiltonian $H_{\text{ave}}$; the second term describes a damping dynamics with the operator $\mathcal{D}$ determined only by the second order noise correlations. As an illustration, consider $H[X(t)]=H_{\text{ave}}+X(t)H_{1}$ [thus $A[X(t)]=A_{\textrm{ave}}+X(t)A_1$]. For the example of noises $X(t)$ with exponential correlations, $\langle X\rangle _{s}=0$ and $\langle X(\tau),X(0)\rangle_{s}=\sigma^{2}\exp(-|\tau|/\tau _{c})$, we have $\mathcal{D}=\sigma^{2}\tau
_{c}A_{1}A_{1}$ for $t\gg\tau_c$ when Eq. (\ref{Rho0t1}) becomes
\begin{equation}
\dot{\rho}_{s}=-i[H_{\text{ave}},\rho _{s}]-\sigma ^{2}\tau
_{c}[H_{1},[H_{1},\rho _{s}]].\nonumber
\end{equation}
Similar equation arises in the main text in the concrete example of a fast two-state telegraph noise [see Eq. (\ref{Telefast})].

\textit{Quasi-static limit}-- We now turn to the quasi-static case, when the relevant times (say, time for experimental observation) are much shorter than the noise correlation time, $t\ll\tau_c$. For this time the noise distribution is effectively frozen to the initial one, and hence we ignore $\mathcal{L}$ in Eq. (\ref{Master_density}) when it reduces
to 
\begin{equation}
\dot{\rho}(X,t)=-i[H(X),\rho(X,t)]\equiv A(X)\rho(X,t).  \label{Static}
\end{equation}
Given an initial $\rho(X,0)$, solution of Eq. (\ref{Static}) gives the average density matrix for times $t\ll\tau_c$ as $\rho_s(t)=
\int dX\exp[A(X)t]\rho(X,0)$ (valid to lowest order of $\tau_c^{-1}$). 

\section{A lattice model for colored Gaussian noise}

\label{App:gaussian}

Here we present a lattice model for colored Gaussian noise $X(t)$
(Ornstein-Uhlenbeck process \cite{Gardiner2010Stochastic}), characterized by a mean value $\langle X(t)\rangle
_{s}$ and a variance $\langle X(t+\tau ),X(t)\rangle _{s}=\sigma ^{2}\exp (-|\tau
|/\tau _{c})$. The basic idea is to form a multistate noise with $N_r$ \textit{
independent} two-state telegraph noises: $X(t)=\sum_{r=1}^{N_r}X_r(t)$. Each telegraph $X_r(t)$ flips between $a^{\prime }$ and $
b^{\prime }$ at a rate $\kappa$, with $\langle X_r\rangle_s=(a^{\prime }+b^{\prime })/2$ and $\left\langle X_r(t+\tau),X_{r'}(t) \right\rangle_s=\delta_{r,r'}\sigma^{\prime
2}\exp(-\tau|/\tau_c)$. Here $\sigma^{\prime 2}=(a^{\prime }-b^\prime)^{2}/4$
and $\tau_c=1/2\kappa$ as in the main text. Thus by construction we have $\langle X(t)\rangle_s=N_r\langle X_r\rangle_s$, and $\sigma^2=N_r{
\sigma^{\prime }}^2$. The noise $X(t)$ can be viewed as resulting from many independent two-level fluctuators with the same jump rate, so that the
instantaneous value of $X(t)$ switches randomly among $N_{r}+1$ discrete values 
$\{X_m\}$ $(m=0,..,N_r)$ with
\begin{equation}
X_m=ma^{\prime }+(N_r-m)b^{\prime }.  \label{Xm}
\end{equation}
We remark that, for \textit{given} $\langle X(t)\rangle_s$ and $\sigma$ of the noise $X(t)$, the values $a^{\prime}$ and $b^{\prime}$ of each telegraph
are determined from scaling relations: $a^{\prime }=\langle
X(t)\rangle_s/N_r-\sigma/\sqrt{N_r}$ and $b^{\prime }=\langle
X(t)\rangle_s/N_r+\sigma/\sqrt{N_r}$.

The stochastic property of the above noise $X(t)$ is described by a probability density $P(X_m,t)$ for finding $X(t)=X_m$ at time $t$, i.e.
\begin{equation}
P(X_m, t)=\frac{m!}{(N_r-m)!N_r!}\left[P(a^{\prime },t)\right]^m\left[
P(b^{\prime },t)\right]^{N_r-m}. \label{Pmt}
\end{equation}
Here $P(a^{\prime }/b^{\prime },t)$ is the probability for a telegraph being in state $a^{\prime
}(b^{\prime })$ at time $t$ with its evolution given by Eq.~(\ref{Telegraph}). Hence we obtain
\begin{equation}
\partial_t P(X_m,t)=\sum_{n=0}^{N_r}\mathcal{L}_{mn}P(X_n,t),  \label{Ljump}
\end{equation}
with $\mathcal{L}_{m,m}=-\kappa N_r$, $\mathcal{L}_{m,m+1}=\kappa\left(m+1
\right)$, $\mathcal{L}_{m,m-1}=\kappa\left(N_r-m+1\right)$, and $\mathcal{L}
_{m,n}=0$ for $n\neq m,m\pm 1$. 

For large $N_r$, the distribution (\ref{Pmt}) approaches a Gaussian distribution as ensured by the central limit theorem \cite{Gardiner2010Stochastic}, and Eq. (\ref{Ljump})
represents the Fokker-Planck equation for an Ornstein-Uhlenbeck process \cite{Gardiner2010Stochastic}. To see this,
we note that $X_{m+1}-X_m=b^{\prime }-a^{\prime }
\sim 1/\sqrt{N_r}$ (with fixed $\sigma$ of the noise), so that when $N_r\rightarrow \infty$ we can replace $X_m$ with the continuous variable $X$ and expand $
P(X_{m\pm 1})$ in terms of $\Delta X=b^{\prime }-a^{\prime
}$ as $P(X\pm\Delta X)\approx P(X,t)\pm (b^{\prime}-a^{\prime})\partial_XP(X,t)
+\frac{(b^{\prime}-a^{\prime})^ 2}{2}\partial_X^2P(X,t)$. In view of $m=(N_rb^{\prime}-X)/(b^{\prime }-a^{\prime})$ from Eq.~(\ref{Xm}), we
obtain from Eq.~(\ref{Ljump}) that
\begin{equation}
\partial_tP(X,t)=\left[2\kappa\frac{\partial}{\partial X}\left(X-\langle
X\rangle_s\right)+\frac{1}{2}D\frac{\partial^2}{\partial X^2}\right]P(X,t). 
\notag
\end{equation}
This is the Fokker-Planck equation governing the Ornstein-Uhlenbeck
process \cite{Gardiner2010Stochastic}, where we identify a drift velocity $2\kappa
$, and a diffusion constant $D=4\kappa \sigma^2$.

Thus the quantum dynamics of a system in (discretized) colored Gaussian noise is governed by the generalized master equation of a form (\ref{Lattice}) with $X_m$ and $\mathcal{L}_{mn}$ given by Eqs. (\ref{Xm}) and (\ref
{Ljump}), respectively -- for large $N_r$, it conveniently approximates Eq. (\ref{Master_density}) for a colored Gaussian noise when $\mathcal{L}(X)$ is the Fokker-Planck operator corresponding to an Ornstein-Uhlenbeck process.

\section{Asymptotic long-time Majorana edge correlations in a quenched Kitaev
Chain}

\label{App:quench}

Here we derive the asymptotic Majorana edge-mode correlation at large times ($t\rightarrow +\infty$) after a global quench in the chemical potential of a Kitaev Hamiltonian, with the quench from $\mu_0$ to $\mu_f$. Specifically, we assume the system is initially in the ground state $|0\rangle$ of
Hamiltonian $H(\mu_{0})$ with $|\mu _{0}|<2J$ in the
topological phase, supporting two Majorana edge modes $\gamma_{L/R}$. Suppose
at time $t=0$, the Hamiltonian is globally
quenched from $H(\mu_{0})$ to $H(\mu _{f})$ with $|\mu_{f}|\lesssim 2J$.
Our goal is to derive the asymptotic ($t\rightarrow +\infty$) Majorana edge
correlation 
\begin{equation}
G_{\infty }=\lim_{t\rightarrow+\infty}\Big\langle 0\Big|e^{iH(\mu_f)t}
\left[-i\gamma_L\gamma_R\right]e^{-iH(\mu_f)t}\Big|0\Big\rangle.
\label{Glongtime}
\end{equation}

\begin{figure}[tb]
\centering
\includegraphics[width= 0.78\columnwidth]{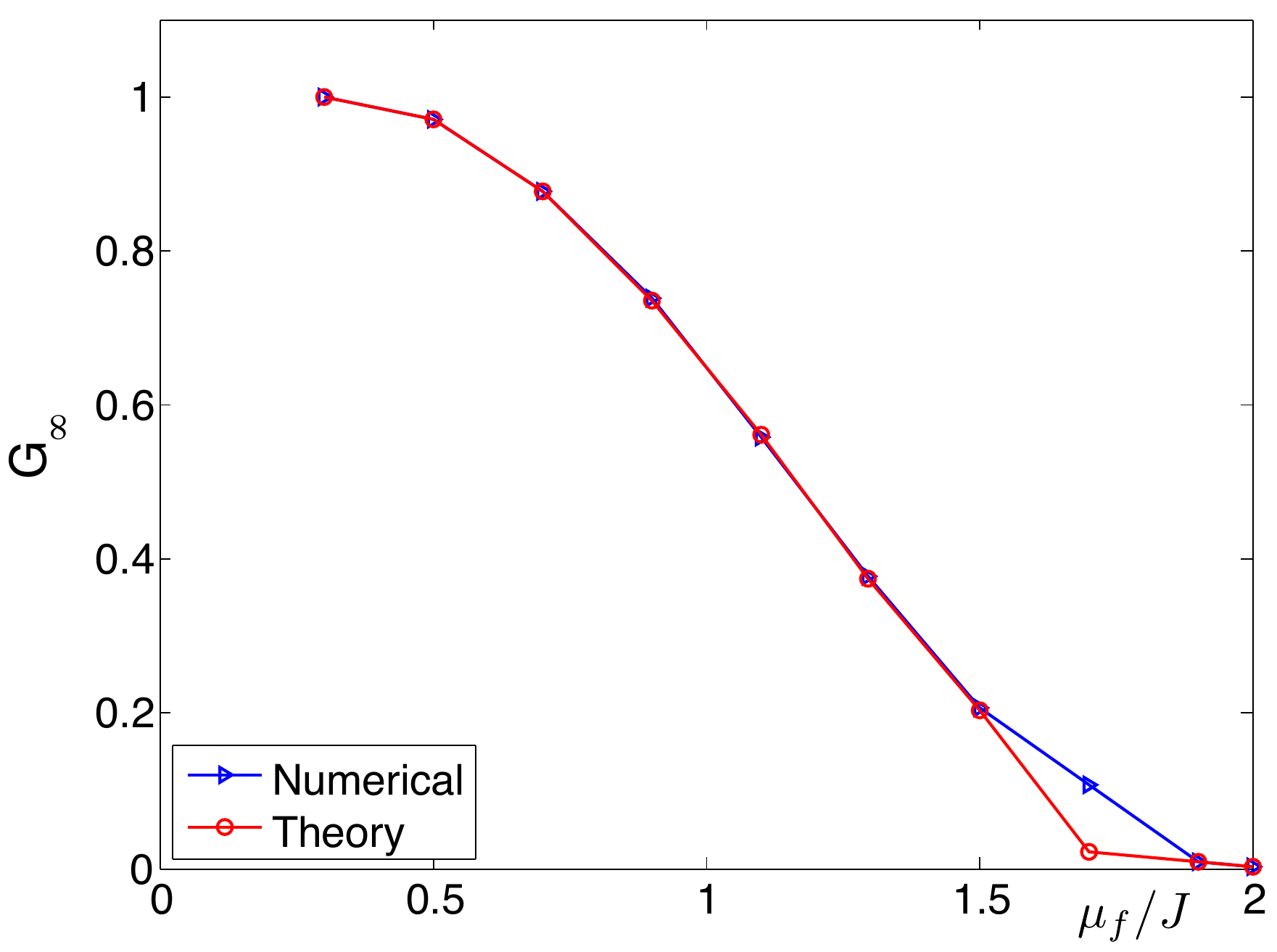}
\caption{{\protect\small {Long-time asymptotic Majorana edge-mode correlation $G_\infty$ (\ref{Glongtime}) as a function of the post-quench chemical potential $\protect\mu_f$, when a Kitaev
Hamiltonian is globally quenched from $H(\protect\mu_0)$ to $H(\protect\mu_f)$, with $|\mu_0|,|\mu_f|\lesssim 2J$. Blue
line - numerics, red line - Eq. (\ref{Gf}). We take $\protect\mu
_0=0.3J$, $N=134$, $\Delta=0.72J$. }}}
\label{Fig:edge}
\end{figure}

In the Majorana basis, Eq. (\ref{Glongtime}) can be written as $G_{\infty}=\sum_{i,l=1}^{2N}f^{(0)}_{L,i}f^{(0)}_{R,l}G^\infty_{il}$, with $G^\infty_{il}=\lim_{t\rightarrow+\infty}-i\langle c_{i}c_{l}\rangle$, and $f^{(0)}_{L/R,l}$ describes left (right) Majorana modes of the initial Hamiltonian. As an
illustration, we calculate $G^\infty_{il}$ for $i=2j_{1}-1$ and $l=2j_{2}$
with $j_{1}\sim 1$ (near the left edge) and $j_{2}\sim N$ (near the right
edge) such that $c_{2j_{1}-1}=a_{j_{1}}+a_{j_{1}}^{\dag}$ and $
c_{2j_{2}}=-i(a_{j_{2}}-a_{j_{2}}^{\dag })$. Using Bogoliubov transformation 
$a_{j}=\sum_{\mu }u_{j\mu }\beta _{\mu }+\nu _{j\mu }^{\ast }\beta_{\mu}^{\dag}$, we diagonalize the post-quench Hamiltonian $H(\mu_{f})$
as $H(\mu_{f})=\sum_{\mu}\epsilon_{\mu}\beta_{\mu}^{\dag}\beta_{\mu}
$ (up to unimportant constant) with $\beta _{\nu }$ being the quasi-particle
operators for the Hamiltonian $H(\mu _{f})$. With this we obtain $
e^{iH(\mu _{f})t}a_{j}e^{-iH(\mu_{f})t}=\sum_{\mu}u_{j\mu}\beta_{\mu
}e^{-i\epsilon_{\mu }t}+\nu_{j\mu}^{\ast}\beta_{\mu}^{\dag}e^{i\epsilon_{\mu}t}$, and similarly for $e^{iH(\mu_{f})t}a_{j}^{\dag }e^{-iH(\mu_{f})t}$. After substituting this into the expression for $G^\infty_{il}$, and neglecting oscillating terms (the ones containing $\beta_{\mu}\beta_{\nu}$, $\beta_{\nu}^{\dag}\beta_{\mu}^{\dag}$, and $\beta_{\mu}^{\dag}\beta_{\nu}$ with $\mu\neq\nu$) which are averaged to zero after times larger than inverse band-width, we obtain
\begin{equation}
G^\infty_{2j_{1}-1,2j_{2}}\approx-\sum_{\mu}\left( A_{j_{1}j_{2}\mu
}\langle 0|\beta_{\mu}^{\dagger}\beta _{\mu}|0\rangle
+B_{j_{1}j_{2}\mu}\langle 0|\beta_{\mu}\beta_{\mu }^{\dagger}|0\rangle\right), \nonumber
\end{equation}
with $A_{j_{1}j_{2}\mu }=(u_{j_{1}\mu}^{\ast}+\upsilon_{j_{1}\mu}^{\ast})u_{j_{2}\mu}+(\upsilon_{j_{1}\mu}+u_{j_{1}\mu})u_{j_{2}\mu }^{\ast}$
and $B_{j_{1}j_{2}\mu }=(u_{j_{1}\mu }+\upsilon _{j_{1}\mu })\upsilon_{j_{2}\mu }^{\ast}+(\upsilon_{j_1\mu }^{\ast}+u_{j_1\mu}^{\ast})\upsilon_{j_2\mu }$. To calculate the correlations in the above expression, we use the
second Bogoliubov transformation $a_{j}=\sum_{\sigma}u_{j\sigma}^{(0)}\alpha_{\sigma}+\nu_{j\sigma }^{(0)\ast }\alpha_{\sigma }^{\dag}$
which diagonalizes the initial Hamiltonian as $H(\mu _{0})=\sum_{\sigma}\epsilon_{\sigma}\alpha_{\sigma}^{\dag}\alpha_{\sigma}$ (again up to
unimportant constant) in terms of quasiparticles operators $\alpha_{\sigma}$ ($\alpha_{\sigma}^{\dag}$), such that $\alpha_{\sigma}|0\rangle =0$. Then, by virtue of the relation $\beta_{\mu}=\sum_{\sigma}C_{\mu\sigma}\alpha _{\sigma }+D_{\mu\sigma}^{\ast}\alpha_{\sigma}^{\dag}$, with $C_{\mu\sigma}=\sum_{j}u_{j\mu}^{\ast}u_{j\sigma }^{(0)}+\upsilon_{j\mu}^{\ast}\upsilon_{j\sigma}^{(0)}$ and $D_{\mu\sigma}=\sum_{j}u_{j\mu}\upsilon_{j\sigma}^{(0)}+\upsilon_{j\mu}u_{j\sigma}^{(0)}$, we arrive at 
\begin{equation}
G^\infty_{2j_{1}-1,2j_{2}}=-\sum_{\mu}\Big(A_{j_{1}j_{2}\mu}\sum_{\sigma}|D_{\mu\sigma}|^{2}+B_{j_{1}j_{2}\mu}\sum_{\sigma }|C_{\mu\sigma }|^{2}
\Big).  \label{G2}
\end{equation}

Equation (\ref{G2}) contains contributions from both the edge ($\mu=M$) and
bulk modes ($\mu\neq M$) of $H(\mu_{f})$ ($|\mu _{f}|\lesssim 2J$). Due to
the gapped energy spectrum, the bulk contribution (modes with $\mu\neq M$)
to the correlations between the edges are exponentially suppressed, and therefore can be ignored in Eq. (\ref{G2}) in the thermodynamic
limit. On the other hand, for the edge contribution ($\mu=M$), we use
Majorana wave functions $g_{L/R,j}$ of Hamiltonian $H(\mu_{f})$ in the fermionic representation: $g_{L,j}=u_{jM}+\upsilon_{jM}$ and $g_{R,j}=u_{jM}-\upsilon_{jM}$. Keeping
in mind the localization character of the Majoranas at edges, we have $
A_{j_1j_2M}\approx g_{L,j_1}g_{R,j_2}$ and $B_{j_1j_2M}\approx -g_{L,j_1}g_{R,j_2}$.
Equation (\ref{G2}) is then simplified as $G^\infty_{2j_{1}-1,2j_{2}}\approx
g_{L,j_1}g_{R,j_2}\sum_{\sigma}\Big(|C_{M\sigma}|^{2}-|D_{M\sigma }|^{2}\Big)$. This expression further involves contributions from the edge $\sigma=M$
and bulk modes $\sigma\neq M$ of initial Hamiltonian $H(\mu _{0})$. For
same reasons discussed earlier, we ignore the exponential small bulk
contributions ($\sigma\neq M$). For the rest edge contribution ($\sigma=M$), we write $u_{jM}^{(0)}=(1/2)(g_{L,j}^{(0)}+g_{R,j}^{(0)})$ and $\upsilon
_{jM}^{(0)}=(1/2)(g_{L,j}^{(0)}-g_{R,j}^{(0)})$ with $g_{L/R,j}^{(0)}$ the
initial Majorana wave functions of Hamiltonian $H(\mu _{0})$ in the fermionic basis. Thus by using $
C_{MM}=\frac{1}{2}\sum_{j}\Big(g_{L,j}g_{L,j}^{(0)}+g_{R,j}g_{R,j}^{(0)}\Big)
$ and $D_{MM}=\frac{1}{2}\sum_{j}\Big(
g_{L,j}g_{L,j}^{(0)}-g_{R,j}g_{R,j}^{(0)}\Big)$, we find 
\begin{equation}
G^\infty_{2j_{1}-1,2j_{2}}\approx g_{L,j_1}g_{R,j_2}\Big[\sum_{j=1}^Ng_{L,j_1}g_{L,j_1}^{(0)}\Big]\Big[
\sum_{j=1}^Ng_{R,j_2}g_{R,j_2}^{(0)}\Big].  \notag
\end{equation}

Calculations of other components of $G^\infty_{il}$ are similar, and we finally get
\begin{equation}
G_{\infty}=\Big[\sum_{j=1}^Ng_{L,j}g_{L,j}^{(0)}\Big]^{2}\Big[
\sum_{j=1}^N g_{R,j}g_{R,j}^{(0)}\Big]^{2}.  \label{Gf}
\end{equation}
We see that, after the quench, the Majorana edge correlation in the long
time approaches an asymptotic value, which is determined only by the overlap
between the wavefunctions of the Majorana edge modes of $H(\mu _{0})$ and $
H(\mu _{f})$ (i.e. $g_{L/R,j}^{(0)}$ and $g_{L/R,j}$). Figure \ref{Fig:edge} shows numerical results for $G_{\infty}$ (blue curve) as a function of $\mu
_{f}$, which are compared to predictions from Eq.~(\ref{Gf}) (red curve). A
good agreement is clearly found, with deviations only appear near the
critical point when the energy gap becomes small. We thus conclude that
Majorana edge correlations relax to a finite value after a quench within
the topological phase, which decreases with $\mu _{f}$ in the post-quench
Hamiltonian and eventually varnishes for $\mu _{f}=2J$.

\end{document}